\documentclass[11pt]{article}
\usepackage[a4paper, margin=1in]{geometry}

\usepackage[usenames]{color}
\usepackage[figuresright]{rotating}
\usepackage{boxedminipage}
\usepackage[colorlinks,citecolor=blue,urlcolor=blue]{hyperref}
\usepackage{epstopdf}
\usepackage{natbib}
\usepackage{multirow}
\usepackage{enumerate}
\usepackage{enumitem}
\usepackage{verbatim}
\usepackage{bbm}
\usepackage[dvipsnames]{xcolor}
\usepackage{bm}
\usepackage{mathrsfs,color,dsfont}
\usepackage{algorithm}
\usepackage{algorithmic}
\usepackage{extarrows}
\usepackage{subcaption}
\usepackage{booktabs}
\usepackage{array} 
\usepackage{makecell}

\usepackage{amsthm,amsmath,amssymb,amsfonts, amsbsy, mathtools, epsfig}

\newcommand{\blind}{1}

\renewcommand{\baselinestretch}{1.31}
\newcommand{\linsp}{\renewcommand{\baselinestretch}{1.31}}
\newcommand{\linsps}{\renewcommand{\baselinestretch}{1.3}}

\newcommand{\Norm}[1]{\left\Vert #1\right\Vert}			
\newcommand{\vect}[1]{\boldsymbol{#1}}				

\newcommand{\tp}{{\top}}						
\newcommand{\sgn}{{\rm sgn}}						
\newcommand{\diff}{\mathrm{d}}					
\newcommand{\argmin}[1]{\mathop{\rm{argmin}}_{#1}}	
											
\newcommand*\widebar[1]{\,						
	\hbox{%
   		\kern0.01em%
		\vbox{%
			\hrule height 0.5pt		
			\kern0.33ex			
			\hbox{%
				\kern-0.1em		
				\ensuremath{#1}%
				\kern-0.1em		
			}%
		}%
	}%
\,}%
\let\hat\widehat
\let\tilde\widetilde

\newcommand{\mcI}{{\mathcal I}}

\newcommand{\mcL}{{\mathcal L}}					

\newcommand{\mcN}{{\mathcal N}}					

\newcommand{\mbB}{{\mathbb B}}					

\newcommand{\mbE}{{\mathbb E}}					

\newcommand{\mbI}{{\mathbb I}}					

\newcommand{\mbP}{{\mathbb P}}					
\newcommand{\mbR}{{\mathbb R}}					
\newcommand{\mbS}{{\mathbb S}}					

\DeclareMathAlphabet\EuScriptBF{U}{eus}{b}{n}

\newcommand{\supp}{\mathrm{supp}}	

\definecolor{DSgray}{cmyk}{0,1,0,0}

\definecolor{scolor}{cmyk}{0.5,2,0,0}

\newtheorem{theorem}{Theorem}

\newtheorem{proposition}[theorem]{Proposition}
\newtheorem{remark}{Remark}

\newtheorem{condition}{Condition}

\begin{document}


\linsps

\if1\blind
{
	\title{\bf Generalized Rank Regression}
		 \author{Jiyuan Tu,\thanks{School of Statistics and Data Science, Shanghai University of Finance and Economics}\quad Suqi Wu,\thanks{School of Mathematical Science, Shanghai Jiao Tong University}\quad Yichen Zhang,\thanks{Department of Quantitative Methods, Purdue University }\quad and Wen-Xin Zhou\thanks{Department of Information and Decision Sciences, University of Illinois Chicago}
		 \date{}
}		\maketitle
} \fi

\if0\blind
{
	\bigskip
	\bigskip
	\bigskip
	\begin{center}
		{\LARGE\bf Generalized Rank Regression}
	\end{center}
	\medskip
} \fi


\bigskip
\begin{abstract}
Rank regression offers robustness to outliers and heavy-tailed response distributions, invariance to monotonic transformations, and improved efficiency under non-Gaussian errors, making it a versatile tool for analyzing complex data. This paper introduces Generalized Rank Regression (GRR), an extension of classical rank-based methods that accommodates non-monotonic score functions. While aimed at enhancing the statistical efficiency of robust estimators, this generalization results in a potentially non-convex and non-smooth objective function, presenting challenges for both theoretical analysis and algorithmic implementation. We derive a non-asymptotic Bahadur representation of the proposed estimator and establish its asymptotic normality under mild conditions. To address the optimization challenges, we propose a new two-stage sub-gradient descent algorithm that enables efficient computation of GRR estimators with desirable statistical properties. Furthermore, we develop a multiplier bootstrap procedure for conducting statistical inference. A close connection between GRR and variants of quantile regression is uncovered, which demonstrates that GRR and composite quantile regression share asymptotically equivalent variances. The advantages of GRR are illustrated through extensive simulation studies and a real data application. 
\end{abstract}


\linsp


\section{{\large Introduction}}

In classical regression modeling, a common assumption is that the conditional distribution of the response variable, given a set of predictors, is Gaussian, or more generally, sub-Gaussian or sub-exponential. The least squares estimator (LSE), along with many of its variants, is widely used across various fields due to its computational simplicity and favorable statistical properties in both asymptotic and non-asymptotic settings. As a natural extension of the sample mean for mean estimation, the LSE performs well under light-tailed distributions but suffers significant performance degradation when the conditional response distribution is heavy-tailed, a common feature of economic and financial data. This limitation motivates the development of alternative regression methods that are robust to heavy-tailed distributions while retaining asymptotic efficiency comparable to the LSE under normality, or even to the maximum likelihood estimator under a correctly specified model \citep{huber2011robust, zou_yuan.2008aos, wang_peng_etal.2020jasa}. To formalize the problem, consider observed data vectors $\{(\vect{X}_i,Y_i)\}_{i=1}^n$, independently drawn from a linear model:
\begin{equation} \label{eq:lin_model}
Y_i = \vect{X}_i^{\tp}\vect{\beta}^* + \epsilon_i,
\end{equation}
and $\epsilon_i$ denotes a noise term independent of the covariates $\vect{X}_i \in \mathbb{R}^p$. We do not assume that $\epsilon_i$ has zero mean; rather, it represents residual variation in $Y_i$ after adjusting for the covariates. To robustly estimate $\vect{\beta}^*$ against the tail behavior of $\epsilon_i$, alternative methods employing robust loss functions, such as the Huber loss, Tukey's bisquare, and biweight loss functions, have been proposed as substitutes for conventional least squares regression \citep{huber1973robust, beaton1974fitting}.

The aforementioned approaches improve robustness by directly replacing the squared loss with carefully designed alternatives that grow more slowly as the residual magnitude increases. From a different perspective, rank regression, dating back to \cite{jaeckel.1972aoms}, provides a robust alternative by employing a loss function based on linear combinations of ordered residuals, rooted in the theory of linear rank statistics \citep{Vaart.1998, sidak1999theory, hettmansperger2010robust}. This method has found wide-ranging applications, including survival analysis \citep{prentice.1978biom, lai_ying.1992jmva} and independence testing \citep{wang_liu_etal.2024joe}. In the context of regression analysis, linear rank statistics were first employed by \cite{jureckova.1969aoms, jureckova.1971aoms}, and later extended by \cite{jaeckel.1972aoms} through the introduction of Jaeckel's dispersion function. When combined with Wilcoxon scores, this formulation, commonly referred to as rank regression, was further explored by \cite{wang_li.2009biometrics} and \cite{leng.2010sinica}. More recently, \cite{wang_peng_etal.2020jasa} established the tuning-free property of $\ell_1$-regularized rank regression and advocated solving it via linear programming. To improve computational efficiency, \cite{zhou_wang_zou.2023jasa} proposed a convolution-smoothed rank regression method using a local linear approximation algorithm with computational complexity $O(n^2T)$, where $T$ denotes the number of iterations and $n$ is the sample size. This tuning-free property was further extended to the low-rank matrix estimation problem by \cite{cui_shi_etal.jmlr2023}.

\begin{figure}[H]\hspace{3em}
\begin{subfigure}[t]{0.39\textwidth}
\includegraphics[width=\linewidth]{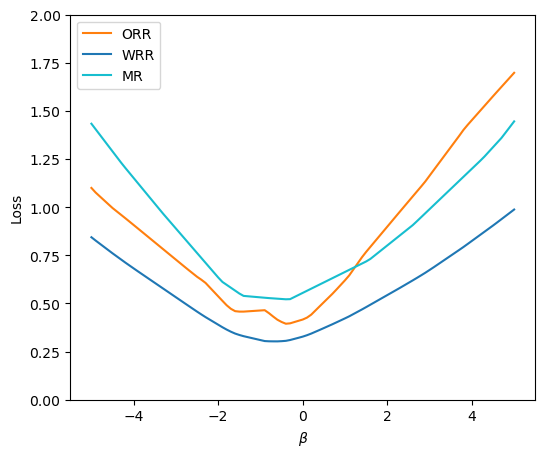}
\end{subfigure}\hfill
\begin{subfigure}[t]{0.39\textwidth}
\includegraphics[width=\linewidth]{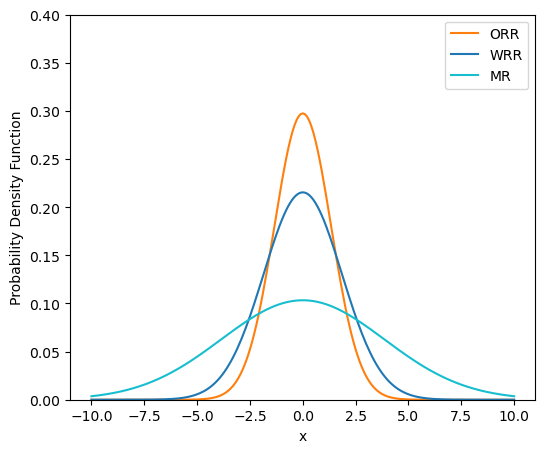}
\end{subfigure}\hspace{5em}

\caption{Empirical loss functions (left) and asymptotic normal density functions (right) for median regression (MR), rank regression with Wilcoxon score (WRR), and rank regression with the optimal score (ORR). The empirical loss is based on $n=10$ data points generated from a univariate linear model with error $\epsilon \sim 0.5 \mathcal{N}(-1.5,1) + 0.5 \mathcal{N}(1.5,1)$.}
\label{fig:nonconvex_loss}
\end{figure}

While canonical rank regression with the Wilcoxon score offers superior robustness relative to least squares regression, it can suffer substantial losses in statistical efficiency under general noise distributions. This is illustrated numerically in Figure \ref{fig:nonconvex_loss}, where the asymptotic variance of the WRR estimator is considerably larger than that of GRR with an optimally specified score function\footnote{Note that in this context, the term ``score function" refers to a function used to assign weights to the ranks of the residuals, which differs from its standard definition as the derivative of the log-likelihood. A theoretical connection between these notions can be found on \citet[page 179]{Vaart.1998}.}. 
To improve efficiency while preserving robustness to heavy-tailed noise, we introduce a unified framework termed Generalized Rank Regression (GRR). This framework extends classical rank regression by accommodating a broader class of score functions. Specifically, we consider a robust estimator defined as
\begin{equation}
\label{eq:beta_hat_intro}
    \hat{\vect{\beta}} =\argmin{\vect{\beta}} \sum_{i=1}^na_n(R_i) \cdot (Y_{i}-\vect{X}_{i}^{\tp}\vect{\beta}) ,
\end{equation}
where $R_i$ denotes the rank of the $i$-th residual $Y_{i}-\vect{X}_{i}^{\tp}\vect{\beta}$ within the set $\{Y_{i}-\vect{X}_{i}^{\tp}\vect{\beta}\}_{i=1}^n$, and $a_n(\cdot)$ is a general score function mapping ranks to real values. Section \ref{sec:grr} provides specific examples of such score functions, including generalizations of those used in classical rank regression, such as the Wilcoxon score. Figure~\ref{fig:nonconvex_loss} also depicts the loss landscapes induced by various score functions. Notably, the loss function derived from the optimal score under a mixture normal error distribution is both non-convex and non-smooth. This example highlights the computational and theoretical challenges introduced by GRR.

Theoretically, we first establish the asymptotic normality of the GRR estimator in its most general form, derived from a Bahadur representation whose remainder term depends subtly on the score function. We also identify the optimal score function $a_n(\cdot)$ that minimizes the estimator's asymptotic variance. It is important to note that the loss function in \eqref{eq:beta_hat_intro} exhibits more intricate behavior than those arising in many classical regression methods, which complicates statistical analysis for two primary reasons:
\begin{enumerate}
    \item[(a)] The loss function in rank regression depends on the residual ranks, $\{R_i\}_{i=1}^n$, which can change abruptly as residuals shift in order. Although the objective function is continuous, it is not differentiable--small changes in the residuals can lead to significant changes in rank, resulting in a fundamentally \emph{non-smooth} objective.

    \item[(b)] More importantly, the generalization of classical rank regression introduces \emph{non-convexity} into the loss landscape. As a result, our theoretical guarantees (consistency and asymptotic normality) are established for stationary points of the objective function in \eqref{eq:beta_hat_intro}. 
\end{enumerate}

The non-smoothness and non-convexity of the GRR loss function not only complicate the theoretical analysis of the estimator's statistical properties but also pose significant challenges for practical computation. While our earlier analysis establishes theoretical guarantees for stationary points, efficiently locating such solutions in practice is equally critical. To this end, we propose a two-stage sub-gradient descent algorithm designed to accommodate the non-convex nature of the GRR objective. In the first stage, we minimize a convex surrogate of the GRR loss using sub-gradient descent with a decreasing step size and a monotone score function. This ensures convergence to a neighborhood of the true solution.  In the second stage, we apply sub-gradient descent to the original (potentially non-convex) GRR loss using a constant step size. This stage leverages Clarke subdifferentials to navigate the non-smooth loss landscape and is initialized with the output from the first stage, resulting in a final estimator with desirable statistical properties. Our tailored step-size schemes for each phase are pivotal, creating a distinct phase transition in the algorithmic convergence behavior; this transition is not merely advantageous but essential for achieving robust convergence. Computationally, our method offers notable improvements. Unlike existing methods for rank regression \citep{wang_peng_etal.2020jasa, zhou_wang_zou.2023jasa}, which incur a per-iteration cost of $O(n^2)$, our algorithm reduces this complexity to $O(n \log n)$---a compellingly small number of iterations for this challenging non-smooth, non-convex problem. The geometric convergence achieved in the second stage of our method guarantees the efficiency of the algorithm.

As shown in Theorem \ref{thm:jd_norm}, the asymptotic variance of the GRR estimator is determined by a complex functional involving both the cumulative distribution function (CDF) and the probability density function (PDF) of the noise variable. Inference based on these asymptotic results necessitates consistent nonparametric estimation of these functions, a process that introduces numerical instabilities and additional tuning parameters. To overcome this, we adopt the weighted/multiplier bootstrap, a resampling technique widely recognized for its effectiveness in estimating standard errors and constructing confidence intervals or regions \citep{efron_tibshirani.1994book, diciccio_efron.1996ss}. We develop a new inference procedure using sub-gradient descent on the bootstrapped loss. By avoiding the exact, potentially time-consuming minimization of each non-convex multiplier bootstrap objective, our method offers both computational efficiency and statistical accuracy for uncertainty quantification. We further establish the theoretical validity of this bootstrap approach.

As a by-product of this study, we uncover a fundamental connection between GRR with general score functions and various forms of quantile regression (QR) \citep{koenker_bassett.1978ecm,zou_yuan.2008aos,jiang_etal.2012sinica}. QR is well-known for its robustness to heavy-tailed noise, with its theoretical properties, such as consistency and asymptotic normality, holding without requiring moment conditions on the noise. To further improve statistical efficiency while preserving robustness, \cite{zou_yuan.2008aos}  introduced equal-weight composite quantile regression (CQR), which achieves higher relative efficiency compared to single-level QR. However, an efficiency gap still remains. Motivated by the shared robustness of rank regression and QR, along with the observation that quantiles are defined via order statistics, we reveal an intrinsic connection between the two frameworks. Several works, including  \cite{wang_li.2009biometrics}, \cite{wang_yu_etal.2019jmlr}, and \cite{wang_peng_etal.2020jasa}, have noted that the classical rank regression and CQR exhibit comparable statistical efficiencies. It has been widely conjectured that rank regression, by implicitly aggregating over an infinite number of quantile levels, mimics the behavior of CQR. Yet, a precise characterization of this relationship is lacking. A key insight of our study is that the population version of the weighted CQR estimator for the $\tau$-th quantile of $\epsilon_i$, denoted by $b_{\tau}^{*}$, is asymptotically equivalent to the $\tau$-th quantile of the residuals $\{Y_i-\vect{X}^{\tp}_i\vect{\beta}\}_{i=1}^n$. This equivalence enables a reformulation of the composite quantile loss into a rank regression loss via term rearrangement. Importantly, this formulation accommodates potentially negative weights in CQR, thereby aligning with the intrinsically non-convex nature of the rank regression objective, an aspect not captured in the original formulation by \cite{jaeckel.1972aoms}.

The rest of the paper is organized as follows. Section \ref{sec:grr} introduces the generalized rank regression and provides illustrative examples. Section \ref{sec:grr_theory} presents a comprehensive theoretical analysis of the statistical properties of GRR. Section \ref{sec:wqr_trans} explores the connection between GRR and composite quantile regression. Section \ref{sec:jd_gd} describes the proposed two-stage sub-gradient descent algorithm for solving GRR. Section \ref{sec:bootstrap} details the multiplier bootstrap procedure developed for conducting statistical inference on GRR estimators. Section \ref{sec:sim} reports numerical results that support our theoretical findings. Python code and data for reproducing our examples are available at \url{https://github.com/suqiwu/GRR_github_ver}.

Throughout the paper, we adopt the following notations. For a vector $\vect{v}=(v_1,v_2,\dots,v_p)^\top$, we define its norms as $|\vect{v}|_1=\sum_{l=1}^p|v_l|$, $|\vect{v}|_2=\sqrt{\sum_{l=1}^pv_l^2}$ and $|\vect{v}|_{\infty}=\sup_{1\leq l\leq p}|v_l|$. The support of $\vect{v}$ is denoted by $\supp(\vect{v})=\{1\leq l\leq p\mid v_l\neq 0\}$. For a matrix $\vect{A}\in \mathbb{R}^{n\times p}$, we define $\|\vect{A}\| = \sup_{\vect{v}\in\mbR^p}|\vect{A}\vect{v}|_2$ and $\Norm{\vect{A}}_{1}:=\max_{1\leq j\leq p}\sum_{i=1}^{n}|a_{ij}|, \Norm{\vect{A}}_{\infty}:=\max_{1\leq i\leq n}\sum_{j=1}^{p}|a_{ij}|$ as various matrix norms. The largest and smallest eigenvalues of $\vect{A}$ are denoted by $\Lambda_{\max}(\vect{A})$ and $\Lambda_{\min}(\vect{A})$, respectively. We denote the indicator function as $\mbI(\cdot)$. For simplicity, we use $\mbS_r^{p-1}(\vect{v})$ and $\mbB^p_r(\vect{v})$ to represent the sphere and the closed ball of radius $r$ centered at $\vect{v}\in\mbR^p$, respectively. For two sequences $a_n$ and $b_n$, $a_n = \Omega(b_n)$ means $b_n=O(a_n)$. Lastly, all generic constants denoted by $C, C_1, C_2, \dots$ are assumed to be independent of $n$ and $p$, and their values may change from instance to instance.

\section{{\large Generalized Rank Regression}}
\label{sec:grr}

Given independent data vectors $\{(\vect{X}_i, Y_i)\}_{i=1}^n$ drawn from the linear model \eqref{eq:lin_model}, we define the generalized rank regression estimator as
\begin{equation}    \label{eq:beta_hat_def}
\begin{aligned}
    \hat{\vect{\beta}} =&\argmin{\vect{\beta} \in \mathbb{R}^p }\mcL(\vect{\beta}) ~~\mbox{ with }~~  \mcL(\vect{\beta}) \equiv   \mcL_n(\vect{\beta}) := \sum_{i=1}^na_n(i)\cdot \{ Y_{(i)}-\vect{X}_{(i)}^{\tp}\vect{\beta} \} ,
\end{aligned}
\end{equation}
where $a_n(i) \in \mathbb{R}$ denotes a real-valued score function for $i=1,2,\dots, n$, and $\{Y_{(i)}-\vect{X}_{(i)}^{\tp}\vect{\beta}\}_{i=1}^n$ represents the order statistics of $\{Y_i-\vect{X}_i^{\tp}\vect{\beta}\}_{i=1}^n$ in ascending order.  Equivalently, the empirical loss function $\mcL(\vect{\beta})$ can be written as $\mcL(\vect{\beta})=\sum_{i=1}^n a_n(R_i)(Y_{i}-\vect{X}_{i}^{\tp}\vect{\beta})$, where $R_i$ denotes the rank of the $i$-th residual $Y_{i}-\vect{X}_{i}^{\tp}\vect{\beta}$ within the set $\{Y_{i}-\vect{X}_{i}^{\tp}\vect{\beta}\}_{i=1}^n$.

The GRR estimator \eqref{eq:beta_hat_def} extends beyond the classical rank regression estimator and encompasses a wide range of well-known regression estimators. We provide several examples of score functions $a_n$ that correspond to some common estimators as special cases of GRR.
\begin{enumerate}[leftmargin=2.4cm]
\item[Example 1](Wilcoxon Score, WRR). Let $a_n(i)=\frac{i}{n+1}-\frac12$. 
\item[Example 2](Sign Score) Let $a_n(i)=\sgn( \frac{i}{n+1} - \frac{1}{2} )$.
\item[Example~3](Single-Level~Score, SRR). Let $a_n(i)\hspace{-0.2em}=\hspace{-0.2em}\tau-\mbI(i<k_0)$,~$k_0\hspace{-0.2em}=\hspace{-0.2em}\arg\min_{k\in\mathbb{N}} |\frac{k}{n}\hspace{-0.2em}-\hspace{-0.2em}\tau|$. 
\item[Example 4](Rank Regression with Sinusoidal Scores). Let $a_n(i)=\sin\{ (\frac{2i}{n+1}-1 )\pi \}$.
\end{enumerate}

The GRR estimator with the classical Wilcoxon score in Example 1 has received much recent interest \citep{wang_peng_etal.2020jasa, cui_shi_etal.jmlr2023}. Examples 2 and 3 exhibit two score functions that attain discrete values. As we demonstrate in Section \ref{sec:wqr_trans}, these estimators are closely related to single-level quantile regressions. Example 4 features a non-monotonic score function resulting in a non-convex loss. 
Importantly, this specific score function is optimal under Cauchy noise, achieving the minimal asymptotic variance. A detailed discussion of the examples is relegated to Section B of the supplementary material. 

It is important to emphasize that we do not restrict the score function $a_n$ to be monotone, as is often required in the literature \citep{jaeckel.1972aoms, hettmansperger2010robust}. This consideration not only broadens the scope and potential applications of the proposed method but is also essential because the optimal score function $a_n$ may indeed be non-monotonic, leading to non-convex objectives (e.g., Example 4), as will be theoretically characterized in Proposition \ref{prop:opt_score}. This non-convexity is further illustrated numerically in Figure \ref{fig:nonconvex_loss} in the introduction. To formally characterize the relationship between the convexity of the objective function $\mcL(\vect{\beta})$ and the monotonicity of the score function  $\{a_n(i)\}$, we establish the necessary and sufficient condition for $\mcL(\vect{\beta})$ to be convex in Proposition \ref{prop:non-convex_loss}.

 \begin{proposition}	\label{prop:non-convex_loss}
    In model \eqref{eq:lin_model}, assume that the noise $\epsilon$ and the covariate $\vect{X}$ admit a joint probability density (and hence possess marginal densities), while allowing for potential dependence between them. Then, with probability one, the loss function $\mcL(\vect{\beta})$ in \eqref{eq:beta_hat_def} is convex if and only if the score function $a_n(i)$ is monotonically increasing.
\end{proposition}

Proposition \ref{prop:non-convex_loss} necessitates a dedicated theoretical analysis of a non-convex, non-smooth problem. Accordingly, in the following section, we first explore the statistical properties of the stationary points defined in \eqref{eq:beta_hat_def}. Subsequently, Section \ref{sec:jd_gd} addresses their algorithmic guarantees, focusing on computational feasibility and efficiency.

\subsection{Theoretical analysis}
\label{sec:grr_theory}

Before delving into the theoretical results concerning the stationary points of \eqref{eq:beta_hat_def}, we first outline some technical conditions.

\begin{condition}   \label{cond:cov}
The covariate vector $\vect{X} \in \mathbb{R}^p$ has mean zero, and its one-dimensional projection $\vect{v}^{\tp}\vect{X}$ is a sub-Gaussian random variable for every $\vect{v}\in\mbS^{p-1}$. Additionally, there exists a constant $\rho\in(0,1)$ such that the covariance matrix $\vect{\Sigma}=\mbE[\vect{X}\vect{X}^{\tp}]$ satisfies $\rho\leq\Lambda_{\min}(\vect{\Sigma}) \leq \Lambda_{\max}(\vect{\Sigma})\leq \rho^{-1}$.
\end{condition}

\begin{condition}   \label{cond:pdf}
    The probability density $f$ of the noise $\epsilon$ is positive everywhere and uniformly bounded. Further, $f'(\epsilon)/f(\epsilon)$ is a sub-exponential random variable
    and $[\log\{f(\epsilon)\}]''$ is uniformly bounded, where $g'$ and $g''$ denote the first- and second-order derivatives of the function $g$, respectively.
\end{condition}

\begin{condition}  \label{cond:score}
    There exists a score-generating function $\varphi:[0,1]\rightarrow\mbR$ such that
    \begin{equation*}
    \begin{aligned}
        &\int_0^1\varphi(x)\diff x=0,~ \int_0^1\varphi^2(x)\diff x=1,~ |\varphi(x)|\leq C_{\varphi},~ c_H := -\int_{-\infty}^{\infty}\varphi\{F(x)\}f'(x)\diff x   >0 ,
    \end{aligned}
    \end{equation*}
    where $F(x)$ denotes the cumulative distribution function of $\epsilon$. With proper rescaling, assume $\max_{1\leq i\leq n} |a_n(i)-\varphi(i/n)|\leq C_a/n$ for some $C_a>0$. Moreover, one of the following holds:
    \begin{itemize}
        \item[3a)] There exists a constant $C_L$ such that $                |\varphi(x)-\varphi(x')|\leq C_L|x-x'|$ for $x,x'\in[0,1]$; 
        \item[3b)] The function $\varphi(x)$ is continuous except at a finite number of points $\{x_1,x_2,\dots,x_K\}$. Denote $x_0=0$ and $x_{K+1}=1$. Then there exists a constant $C_L$ such that $|\varphi(x)-\varphi(x')|\leq C_L|x-x'|$ for $x,x'\in(x_i,x_{i+1})$, $i=0,1,\dots,K.$    
        \end{itemize}
\end{condition}

Condition \ref{cond:cov} imposes standard regularity conditions on the covariate vector \citep{wang_li.2009biometrics, wang_peng_etal.2020jasa, zhou_wang_zou.2023jasa}. Condition \ref{cond:pdf} imposes certain regularity conditions on the probability density function of $\epsilon$, which are satisfied by many common distributions, such as the normal, Laplace, Student's $t$, and Cauchy distributions, as well as any distribution with uniformly bounded $f'(\epsilon)/f(\epsilon)$ and $f''(\epsilon)/f(\epsilon)$. The moment requirement is weaker than those needed for Huber regression, where a $(1+\delta)$-th moment for some $\delta>0$ is typically required for consistency, and $\delta \geq 1$ is necessary for asymptotic normality \citep{sun_etal.2020jasa}. Condition~\ref{cond:pdf} excludes certain parametric families, such as the Beta and Gamma distributions, for which the contiguity of the associated probability measures may fail. Extending the analysis to accommodate such distributions would likely require distribution-specific proof strategies, rather than the unified approach developed here. We therefore leave this direction for future research. Condition \ref{cond:score} delineates the requirements on the score function $a_n(i)$ by approximating to a score-generating function $\varphi(x)$, defined on $[0,1]$. We do not require $\varphi(x)$ to be non-decreasing, as is needed in Theorem 1 of \cite{jaeckel.1972aoms}. This allows for the potential non-convexity of the loss function in \eqref{eq:beta_hat_def}. Instead, we require $-\int_{-\infty}^{\infty}\varphi\{F(x)\}f'(x)\diff x>0$, which implies the local strong convexity of population loss. Additionally, we examine the score-generating function with regard to its continuity, enabling a more nuanced analysis of the estimator. The distinction between Conditions \ref{cond:score}a) and \ref{cond:score}b) is both fundamental and necessary for two key reasons. First, the higher-order rate of convergence differs under these two conditions, which leads to distinct scaling conditions required for the normal approximation in Theorem \ref{thm:jd_norm} below. Second, the previously listed examples satisfy one of these two conditions. Specifically, Condition \ref{cond:score}a) applies to continuous score-generating functions $\varphi$, as seen in Examples 1 and 4, while Condition \ref{cond:score}b) corresponds to discontinuous $\varphi$ with a finite number of jumps, applicable to Examples 2 and 3.
These conditions accommodate a broader spectrum of score functions beyond Jaeckels's dispersion function \citep{jureckova.1971aoms,jaeckel.1972aoms}, which is restrictively generated as $a_n(i) = \varphi\{i/(n+1)\}$. Moreover, the conditions imposed on the covariates $\vect{X}$ in \cite{jureckova.1971aoms} are difficult to verify, whereas our conditions are straightforward. 
Condition~\ref{cond:score} imposes a piecewise Lipschitz continuity assumption on the score-generating function $\varphi$. While this regularity condition excludes certain score functions, such as those that are optimal for the Gaussian distribution, this restriction stems from the limitations of our current proof techniques rather than an inherent constraint of the GRR methodology. Extending the framework to accommodate unbounded domains or derivatives would require delicate truncation arguments, along with carefully balanced assumptions on the tail behavior of both the score function and the underlying error density. As this technical extension does not yield additional methodological insight, we do not pursue it further in this work.

Under these regularity conditions, we establish the Bahadur representation and asymptotic normality of the GRR estimator. Because the objective function $\mcL(\vect{\beta})$ in \eqref{eq:beta_hat_def} may be non-convex, we define a stationary point as any parameter vector $\vect{\beta} \in \mathbb{R}^p$ satisfying $\vect{0} \in \partial\mcL(\vect{\beta})$, where $\partial\mcL$ denotes the Clarke subdifferential of $\mcL$.

\begin{theorem} \label{thm:jd_norm}
    Assume that Conditions \ref{cond:cov}--\ref{cond:score} hold. Then, there exists a stationary point $\hat{\vect{\beta}}$ satisfying
    \begin{equation*}
    	 \hat{\vect{\beta}}-\vect{\beta}^*=c_H^{-1}\vect{\Sigma}^{-1}\frac{1}{n}\sum_{i=1}^n\varphi(F(\epsilon_i))\vect{X}_i + R_n,
    \end{equation*} 
    where $c_H$ is defined in Condition \ref{cond:score}, and the remainder term $R_n$ is bounded as follows:
    \begin{itemize}
    	\item[a)] Under Condition \ref{cond:score}a), $R_n=O_{\mbP}(p^{3/2}\log^{3/2} (n)/n)$;
	\item[b)] Under Condition \ref{cond:score}b), $R_n=O_{\mbP}((p\log n/n)^{3/4}+p^{3/2}\log^{3/2} (n)/n)$.
    \end{itemize}
    Consequently, if the remainder term satisfies $R_n=o_{\mbP}(1/\sqrt{n})$, then for any sequence of non-zero vectors $v_n\in\mbR^p$, we have
    \begin{equation}
        \frac{\sqrt{n}}{\sigma_{\vect{v}_n}}\vect{v}_n^{\tp}(\hat{\vect{\beta}}-\vect{\beta}^*)\xrightarrow{{\rm d}}\mcN(0,1) ~~\text{ with }~~ \label{eq:sigma_v}
        \sigma_{\vect{v}_n}^2 = \frac{1}{c_H^2}\vect{v}_n^{\tp}\vect{\Sigma}^{-1}\vect{v}_n.
    \end{equation} 
\end{theorem}
Theorem~\ref{thm:jd_norm} reveals that the rate of the remainder term in the Bahadur representation depends on the continuity of $\varphi(x)$. Specifically, in Example 1, where the score function $\varphi(x) = x-1/2$ is Lipschitz continuous, the remainder is of order $O_{\mathbb{P}}(p^{3/2}\log^{3/2} (n)/n)$. 
In contrast, the remainder term in case b) captures a slower rate of $O_{\mathbb{P}}( p^{3/4}\log^{3/4}(n) /n^{ 3/4}+ p^{3/2}\log^{3/2} (n)/n )$, which aligns with those in single-level quantile regression \citep{koenker.2005, pan_zhou.2020ima} and can be viewed as a special case of our method (see Example 3). 
Note that the relationship between $p$ and $n$ differs according to the continuity of $\varphi(x)$. For example, when $\varphi$ allows for discontinuity as assumed in Condition 3b), we require $p^3=o(n/(\log n)^3)$ for case 3b), which aligns with those in single-level quantile regression \citep{koenker.2005, pan_zhou.2020ima} and can be viewed as a special case of GRR (see Example 3). To our knowledge, this is the first result of its kind in rank regression, complete with an explicit remainder term. In the literature, \cite{jaeckel.1972aoms} established the asymptotic normality of the estimator with a Jaeckel's dispersion function under the fixed-$p$ setting. \cite{wang_peng_etal.2020jasa} and \cite{cui_shi_etal.jmlr2023} established convergence rates for Wilcoxon-score rank regression (Example 1) in high dimensions. However, their analyses rely on the theory of $U$-statistics, which does not generalize to GRR. In contrast, our approach is based on a refined analysis of the contiguity between two density functions, yielding a sharper result.

\begin{remark}
    Theorem~\ref{thm:jd_norm} allows the score-generating function to be non-monotone, which may induce non-convexity in the empirical loss function. In such settings, characterizing the global minimizer is computationally intractable. Accordingly, Theorem~\ref{thm:jd_norm} provides a rigorous statistical guarantee for a particular stationary point $\hat{\vect{\beta}}$ lying in a local neighborhood of the true parameter $\vect{\beta}^*$, where the population loss satisfies a local strong convexity condition. To connect this theoretical result with practical implementation, Section~\ref{sec:jd_gd} establishes a corresponding computational guarantee. In particular, we show that such a stationary point, with the desired statistical properties, can be effectively obtained via a two-stage gradient-based algorithm tailored to the non-convex landscape.
\end{remark}

Recall from Theorem \ref{thm:jd_norm} that the asymptotic variance of the GRR depends solely on the score-generating function $\varphi(x)$, rather than the explicit form of the score function $a_n(i)$. Moreover, it involves the constant $c_H = -\int_{-\infty}^{\infty}\varphi\{F(x)\}f'(x)\diff x$, assumed to be positive, where $F(x)$ is the distribution function of $\epsilon$ and $\varphi(x)$ is the score-generating function. It is worth noting that the asymptotic variance is invariant to shifts in the location parameter of the noise distribution. That is, $c_H$ remains unchanged when replacing $F(x)$ by $F(x+b)$ for any constant $b$. The GRR estimator defined in \eqref{eq:beta_hat_def} is also invariant to shifts. On the other hand, an intriguing question arises regarding the identification of the optimal score function that minimizes the asymptotic variance expressed in \eqref{eq:sigma_v}. This issue is addressed by the following proposition for any known distribution function $F(x)$.

\begin{figure}[htp]
\begin{subfigure}[t]{0.33\textwidth}
    \includegraphics[width=\linewidth]{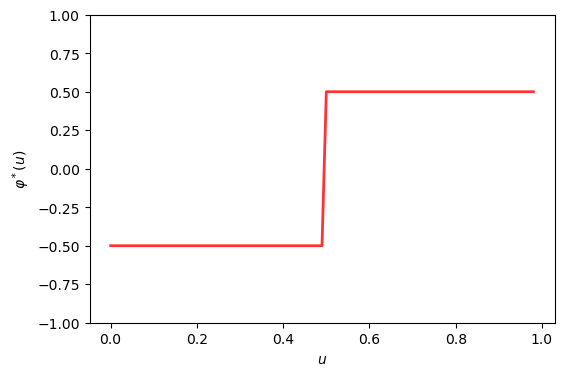}
\end{subfigure}\hfill
\begin{subfigure}[t]{0.33\textwidth}
  \includegraphics[width=\linewidth]{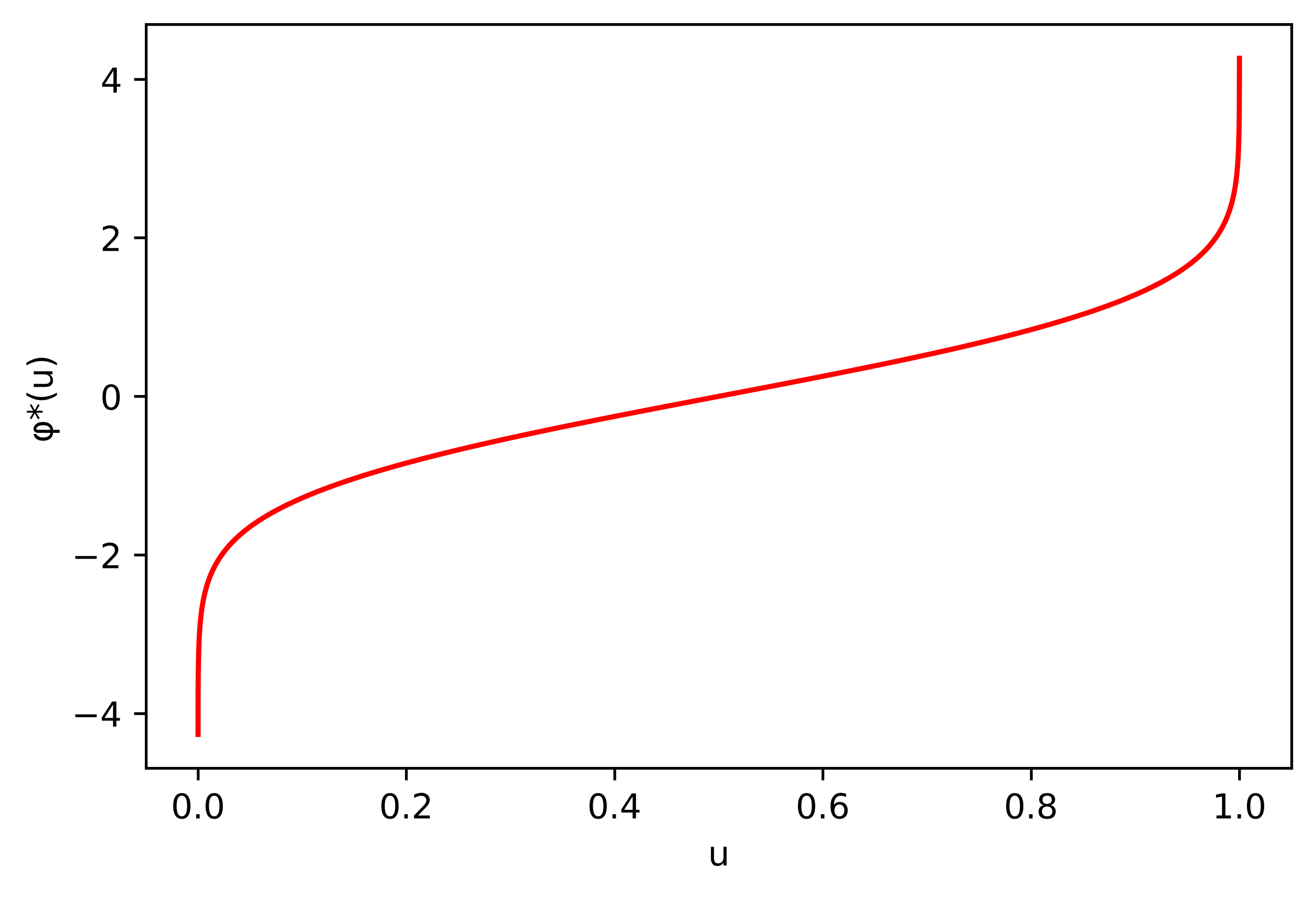}
\end{subfigure}\hfill
\begin{subfigure}[t]{0.33\textwidth}
  \includegraphics[width=\linewidth]{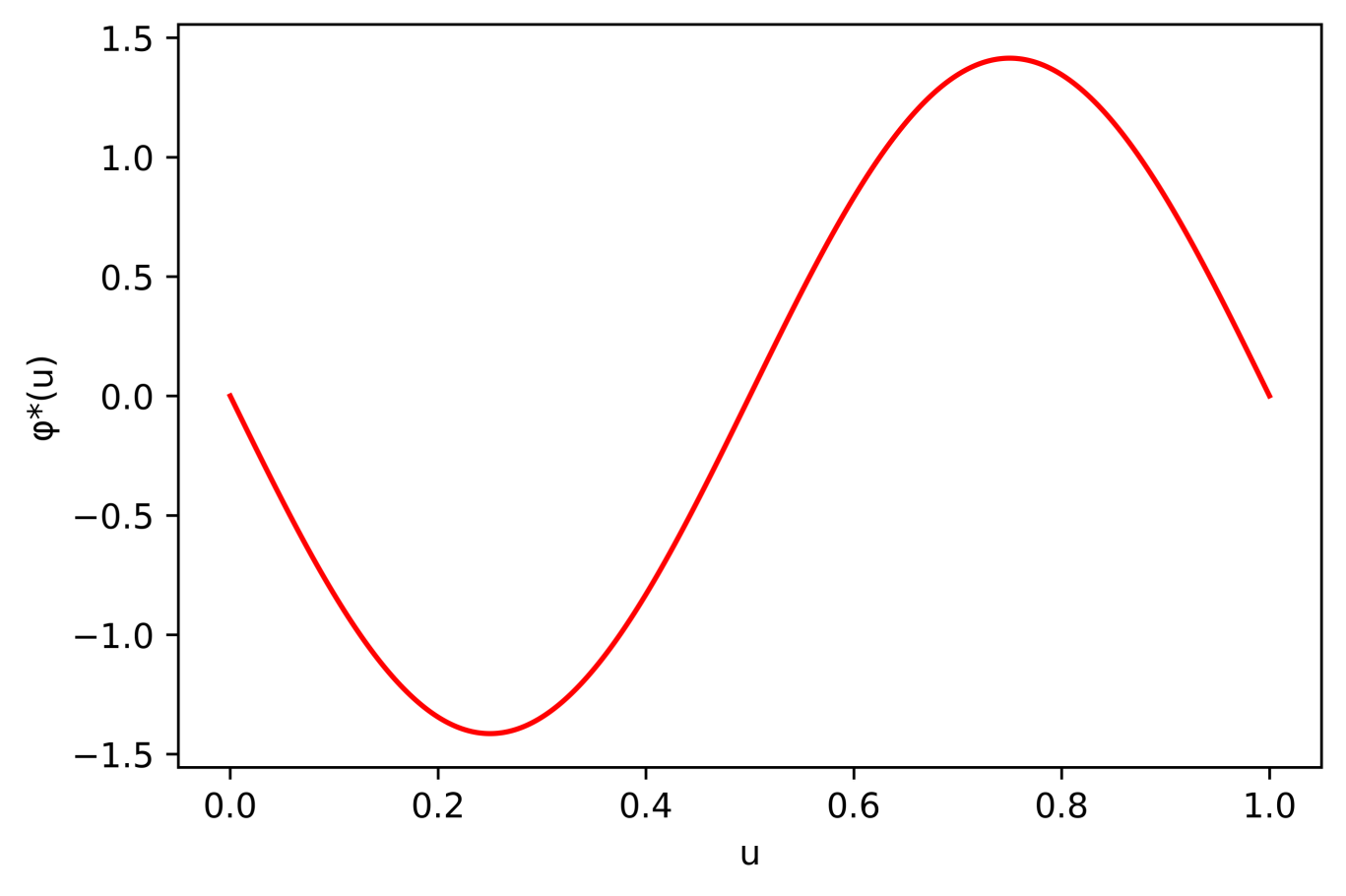}
\end{subfigure}

\caption{Plots of the optimal score-generating functions for the Laplace distribution (left), the standard Normal distribution (middle), and the standard Cauchy distribution (right).}
\label{fig:opt_score}
\end{figure}

\begin{proposition}[Optimal Score Function] \label{prop:opt_score}
The optimal score-generating function $\varphi^*(u)$ that minimizes the asymptotic variance $\sigma^2_{\vect{v}}$ in \eqref{eq:sigma_v} is given by
    \begin{equation}    \label{eq:opt_score}
        \varphi^*(u) = -\frac{1}{\sqrt{\mcI(f)}}\frac{f'\{F^{-1}(u)\}}{f\{F^{-1}(u)\}},\quad u\in[0,1],
    \end{equation}
    where $\mcI(f) = \mbE\{ f'(\epsilon)/f(\epsilon) \}^2$ denotes the Fisher information. In this case, the asymptotic variance is equal to
    \begin{equation*}
        \sigma_{\vect{v}}^2 = \frac{1}{\mcI(f)}\vect{v}^{\tp}\vect{\Sigma}^{-1}\vect{v}.
    \end{equation*}
\end{proposition}

The optimality condition outlined in \eqref{eq:opt_score} aligns with classical findings in linear rank statistics \citep{hettmansperger2010robust}. However, this condition has not been thoroughly explored in the context of GRR estimators. One significant hindrance is that the optimal score $\varphi^*(x)$ may not generally exhibit monotonicity; for instance, the optimal score for the Cauchy distribution is illustrated in Figure \ref{fig:opt_score}. Consequently, this introduces non-convexity into the loss function, a facet not addressed by the theory in \cite{jaeckel.1972aoms}. In this study, we expand our scope to incorporate non-convexity and subsequently develop associated theories and algorithms that guarantee consistency. Additionally, it is important to note that the optimal score for the Laplace distribution exhibits a discontinuity, necessitating the study of GRR under Condition \ref{cond:score}b). This aspect is often overlooked in the rank regression literature.

\begin{remark}\label{rem:ARE}
In light of Proposition \ref{prop:opt_score}, it is evident that an optimally specified GRR can significantly outperform WRR. Specifically, consider an example where the noise $\epsilon$ follows a Student's $t_{\nu}$-distribution with $\nu$ degrees of freedom. We establish in Proposition S1 of the supplementary material that the asymptotic relative efficiency ($\mathrm{ARE}$) of the optimal-score GRR  compared to WRR tends to infinity as $\nu\downarrow 0$. This assertion provides an example where the efficiency of WRR deteriorates to arbitrarily small levels. Consequently, the relative efficiency of optimal-score GRR compared to WRR tends to infinity asymptotically. 
\end{remark}

This highlights the importance of using non-convex loss functions to enhance statistical efficiency. The formal statement and proof are relegated to the supplement.

\begin{remark}
Proposition \ref{prop:opt_score} provides an analytic expression for the optimal score-generating function that minimizes the asymptotic variance when the distribution function $F(x)$ is known. In practice, when $F(x)$ is unknown, one can first select a candidate $\varphi$ to fit an initial GRR estimator $\hat{\vect{\beta}}_0$, then derive the optimal score \eqref{eq:opt_score} based on the empirical distribution of the residuals from $\hat{\vect{\beta}}_0$. This data-driven optimal score function can be subsequently used to refit the GRR estimator.
\end{remark}

While efficiency is a key concern, practitioners often prioritize regression functions that are also robust to model misspecification and outlier contamination.

From a robustness perspective, GRR inherently attains this property when its score function is bounded. For instance, Figure \ref{fig:opt_score} shows that the score-generating functions $\varphi^*$ for single-level rank regression (Examples 2--3) and the standard Cauchy distribution (Example 4) are bounded, making the corresponding GRR estimators robust. In contrast, although the score $\varphi^*$ in the middle plot is monotonic, it becomes unbounded at 0 and 1, leading to a non-robust estimator. Although the proposed GRR estimator does not belong to the class of $M$-estimators typically examined in robust statistics, we provide theoretical insights into how GRR achieves robustness through the influence function (see, e.g., \cite{pj1986robust}) and how the choice of score function impacts this robustness.

\begin{figure}[htp]
\begin{subfigure}[t]{0.33\textwidth}
    \includegraphics[width=\linewidth]{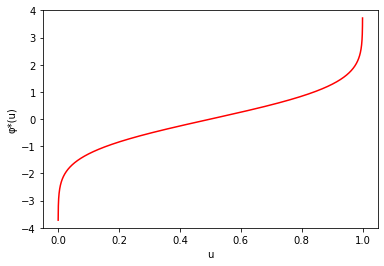}
\end{subfigure}\hfill
\begin{subfigure}[t]{0.33\textwidth}
  \includegraphics[width=\linewidth]{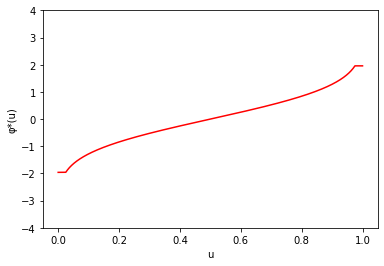}
\end{subfigure}\hfill
\begin{subfigure}[t]{0.33\textwidth}
  \includegraphics[width=\linewidth]{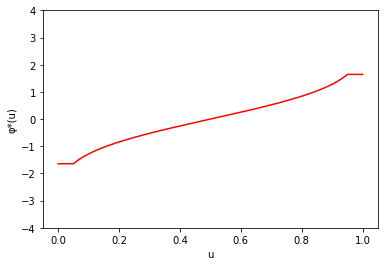}
\end{subfigure}
\caption{Score-generating functions $\varphi$ restricted to be flat on $[0,\varepsilon]$ and $[1-\varepsilon,1]$, derived from the optimal $\varphi^*$ in \eqref{eq:opt_score} for the standard normal distribution: (left) $\varepsilon=0$; (middle) $\varepsilon=0.025$; (right) $\varepsilon=0.05$, highlighting robustness-efficiency trade-offs in GRR.}
\label{fig:threshold_normal}
\end{figure}

\begin{remark}[Robust Score Functions] \label{rem:robust}
We evaluate the influence function, which quantifies the sensitivity of an estimator to any single point in the sample. 
 Let the functional $T$ be the asymptotic value of an estimator sequence $\{T_n\}$. At the population level, the influence function with respect to a point $(\vect{X}, Y)$ is defined as
\[
{\rm IF}\big\{(\vect{X},Y), T, F_{\vect{\beta}^*}\big\}=\lim_{t\rightarrow 0^+}\frac{T\{(1-t)F_{\vect{\beta}^*}+t\delta_{\vect{X},y}\}-T(F_{\vect{\beta}^*})}{t}, 
\]
where $F_{\vect{\beta}^*}$ denotes the distribution of $(\vect{X},Y)$ under model \eqref{eq:lin_model}. Theorem \ref{thm:jd_norm} (Bahadur representation) offers a heuristic understanding of the influence function for GRR, as 
\begin{equation*}
    {\rm IF}\big\{  (\vect{X},Y), T_{\varphi}, F_{\vect{\beta}^*}\big\} = c_H^{-1}\varphi\big\{F(Y - \vect{X}^{\tp}\vect{\beta}^*)\big\}\vect{\Sigma}^{-1}\vect{X}.
\end{equation*}
This influence function provides a guiding principle for designing the score-generation function $\varphi$ to enhance robustness against gross outliers in $Y$. Specifically, when $\varphi$, or equivalently the score $a_n(i)$, is bounded, the influence function is also bounded, indicating that GRR is robust to noise outliers. Examples 1--4 all adhere to this principle.

Moreover, for GRR with sinusoidal scores (Example 4), the values of $\varphi$ at the two extreme points are zero, i.e., $\varphi(0) = \varphi(1) = 0$. This behavior ensures that the influence function approaches zero as $|\epsilon| \rightarrow \infty$, thereby enhancing robustness against significant outliers. It is evident that achieving this level of robustness necessitates that $\varphi$ be non-monotonic, which consequently results in a non-convex objective function. Furthermore, if $\varphi(u) = 0$ for $u \leq u_0$ and $u \geq 1 - u_0$, a portion of extreme observations is effectively discarded as outliers, leading to trimmed and Winsorized regression methods. Although trimming and Winsorization techniques are widely used as robust summary statistics in exploratory data analysis, their regression counterparts are less prevalent, partly due to the non-convexity of the objective function, which complicates both theoretical analysis and computational implementation. 

When the score-generating function $\varphi$ is flat on $[0,x_0]$ and $[x_1,1]$, the breakdown point \citep{donoho1983festschrift} of GRR is asymptotically $\min\{x_0,1-x_1\}$; see Examples 2 and 3. According to the influence function described earlier, an outlier in $Y$ affects the GRR estimator only through the quantity $\varphi(F(\epsilon))$. Consequently, when $\varphi$ is flat near $0$ and $1$, the estimator is asymptotically insensitive to extreme deviations in certain observations. 
In the context of an $\varepsilon$-contamination model, where $Y|\vect{X}\sim(1-\varepsilon)P+\varepsilon Q$, one can further restrict the score-generating function $\varphi$ to be flat outside the interval $[\varepsilon,1-\varepsilon]$ and optimize $\varphi$ within this range to achieve efficiency under $P$. This approach ensures that GRR robustness against contamination from the outlier distribution $Q$, while preserving high efficiency under the target distribution $P$. This highlights the flexibility of GRR in balancing the contamination robustness and statistical efficiency through the appropriate choices of score-generating functions. Figure \ref{fig:threshold_normal} illustrates how such a balance can be achieved by designing $\varphi$ accordingly.
\end{remark}

\subsection{Connections to quantile-based regressions}
\label{sec:wqr_trans}

It is important to note that the asymptotic distribution of GRR, as established in Proposition \ref{prop:opt_score}, is closely related to those of QR and SRR. In this section, we explore the intrinsic connections between GRR and these two methods. We begin by establishing a connection between QR (\citealp{koenker_bassett.1978ecm}) and SRR, which is a special case of GRR as discussed in Example 3.

\begin{remark}[Connection to Quantile Regression]
\label{rem:connection_qr}

Example 3 illustrates a special case of GRR known as single-level rank regression, where the function $a_n(i)$ is given by $\frac{k_0-1}{n}-\mbI\{i<k_0\}$ for an integer $1\leq k_0\leq n$. Let $\tau=\frac{k_0-1}{n}$. According to Condition \ref{cond:score}, the corresponding score-generating function is given by $\varphi(x)=\{\tau-\mbI(x\leq \tau)\}/\sqrt{\tau(1-\tau)}$. Theorem \ref{thm:jd_norm} characterizes its asymptotic covariance matrix as
\begin{equation}\label{eq:asym_cov_SRR}
\vect{V}^{\rm (SRR)}:= c_H^{-2} \vect{\Sigma}^{-1} ~~\mbox{ with }~~ c_H=-\int_{-\infty}^{\infty}\varphi\{F(x)\}f'(x)\diff x=
\frac{f\{F^{-1}(\tau)\}}{\sqrt{\tau(1-\tau)}}.
\end{equation}
The asymptotic covariance matrix of SRR is thus identical to that of quantile regression, where $f(\cdot)$ and $F(\cdot)$ are the probability density and cumulative distribution functions of the noise, respectively. In the following, we demonstrate the connection between SRR and QR. In a QR model, we assume that the 
$\tau$-th conditional quantile of $Y$ depends linearly on $\vect{X}$, with an intercept $b_{\tau}$, i.e., $Q_{Y}(\tau)=\vect{X}^{\tp}\vect{\beta}_{\tau}+b_{\tau}$. The QR estimator of $(b_{\tau},\vect{\beta}_{\tau})$, denoted by $(\hat b_\tau, \hat{\vect{\beta}}_\tau)$, minimizes the empirical check loss
\begin{equation*}
    \mcL^{\rm (QR)}(b,\vect{\beta}):=\sum_{i=1}^n\rho_{\tau}(Y_i - b-\vect{X}_i^{\tp}\vect{\beta})=\sum_{i=1}^{n}\left[\tau-\mbI\{\epsilon_i(\vect{\beta})\leq b\}\right]\cdot \{ \epsilon_i(\vect{\beta}) - b \} ,
\end{equation*}
where $\epsilon_i(\vect{\beta}) = Y_i - \vect{X}_i^{\tp}\vect{\beta}$. Let $\{\epsilon_{(i)}(\vect{\beta} )\}$ be the order statistics of $\{ \epsilon_i(\vect{\beta})\}$ for each $\vect{\beta}$. Since the ordering of data does not alter the empirical distribution, we can equivalently express the QR objective \eqref{eq:qr} using the ordered residuals, that is, 
\begin{equation}\label{eq:qr}
\mcL^{\rm (QR)}(b,\vect{\beta})=\sum_{i=1}^{n}  [\tau-\mbI\{\epsilon_{(i)}(\vect{\beta})\leq b\} ]\cdot \{ \epsilon_{(i)}(\vect{\beta}) - b  \}.
\end{equation}
Let $\tau =(k_0-1)/n$ for some positive integer $k_0$. For any fixed $\vect{\beta}$, minimizing \eqref{eq:qr} over $b$ leads to a minimizer $\hat{b}_{\tau}$ taking the sample quantile of the residuals $\{\epsilon_{(i)}(\vect{\beta})\}$, and thus $\sum_{i=1}^{n}  (\tau-\mbI\{\epsilon_{(i)}(\vect{\beta})\leq \hat b_\tau \} )\cdot  \hat b_\tau =  0$. Since $\{\epsilon_{(i)}(\vect{\beta})\}$ is ordered, it holds that $\epsilon_{(i)}(\vect{\beta})\leq \hat{b}_\tau $ if and only if $i< k_0$. 
On the other hand, recall the SRR objective
\[
\mcL^{\rm (SRR)}(\vect{\beta})=\sum_{i=1}^{n}a_n(i)\cdot\epsilon_{(i)}(\vect{\beta})=\sum_{i=1}^{n} \{ \tau - \mbI(i<k_0) \} \cdot \epsilon_{(i)}(\vect{\beta}).
\]
The above objective is essentially the same as \eqref{eq:qr}. In light of the above, the asymptotic properties of the QR and SRR estimators should align. Additionally, our simulation studies in Section \ref{sec:sim} confirm that the finite-sample performance of SRR closely matches that of QR.
\end{remark}

As the name suggests, single-level rank regression uses a single-step score function with a discontinuity at $\tau$, and we demonstrate its equivalence to QR. A known drawback of QR is that its relative efficiency with respect to the LSE can be arbitrarily small. To overcome this limitation, \cite{zou_yuan.2008aos} introduced composite quantile regression, which  simultaneously incorporates multiple QR models into a unified objective, given by
\begin{equation}    \label{eq:ecqr}
(\hat{b}_1,\hat{b}_2,\dots,\hat{b}_K,\, \hat{\vect{\beta}}) = \argmin{b_1,b_2,\dots,b_K, \, \vect{\beta}}\;\sum_{k=1}^K \sum_{i=1}^n\rho_{\tau_{k}}(Y_i-b_k-\vect{X}_i^{\tp}\vect{\beta}) ,
\end{equation}
where $K$ quantile levels $0<\tau_1<  \cdots <\tau_K <1$ are predetermined, and $\hat{b}_1,\dots,\hat{b}_K$ are the estimators for the corresponding quantiles $b^*_{\tau_k}=F^{-1}(\tau_k)$. Following a similar derivation as in Remark~\ref{rem:connection_qr}, CQR closely aligns with WRR in Example~1, serving as a special case of GRR. Furthermore, the minimizer of WRR shares the same asymptotic variance as CQR when $K$ approaches infinity \citep{wang_yu_etal.2019jmlr, wang_peng_etal.2020jasa}. While the proposed equal-weight CQR substantially improves statistical efficiency over QR, an efficiency gap remains between CQR and the oracle maximum likelihood estimator. As shown in Proposition~\ref{prop:opt_score}, GRR with an optimally designed score function further improves asymptotic efficiency beyond that of WRR. In the following, we establish that GRR is closely related to a weighted version of CQR, encompassing WRR as a special case while generalizing to arbitrary score functions. Moreover, we show that GRR with non-monotonic score functions corresponds to non-convex CQR, where the non-convexity arises from negative weights. Additionally, GRR and the corresponding CQR share asymptotic variances when $K$ approaches infinity.

\begin{remark}[Connection to Composite Quantile Regression]\label{rem:connect_cqr} 
We show that GRR with a general score function $a_n(i)$ 
is closely connected to a weighted variant of CQR \citep{jiang_etal.2012sinica}, for which the objective function is defined as
\begin{equation}    \label{eq:wcqr}
\mcL^{\rm (CQR)}(b_1,b_2,\dots,b_K,\vect{\beta})=\sum_{k=1}^K\sum_{i=1}^n w_k\rho_{\tau_{k}} (Y_i-b_k-\vect{X}_i^{\tp}\vect{\beta} ) .
\end{equation}
The relationship between the weights $\{w_k\}_{k=1}^n$ and the score function $\{a_n(i)\}_{i=1}^n$ will be revealed later. Since ordering does not change the empirical distribution, one can replace $(\vect{X}_i,Y_i)$ in \eqref{eq:wcqr} with $(\vect{X}_{(i)},Y_{(i)})$, where the subscript $(i)$ means that the corresponding residual is the $i$-th smallest among all residuals. As demonstrated in Remark \ref{rem:connection_qr}, by setting $K=n$ and choosing the quantile levels as $\tau_k=k/n$, one can 
replace $b_k$ with $\epsilon_{(k)}$. Consequently, the loss function in \eqref{eq:wcqr} can be approximately reformulated as 
\footnotesize
\begin{align*}
\sum_{k=1}^K\sum_{i=1}^n w_k\rho_{\tau_{k}} ( \epsilon_{(i)}-\epsilon_{(k)} ) =\sum_{i=1}^n\sum_{k=1}^nw_k\big[  \tau_k-\mbI \{ \epsilon_{(i)}\leq\epsilon_{(k)} \} \big]\epsilon_{(i)}-\sum_{k=1}^n\sum_{i=1}^n w_k \big[ \tau_k-\mbI\{ \epsilon_{(i)}\leq\epsilon_{(k)} \} \big] \epsilon_{(k)}.
\end{align*}
\normalsize
Since $\{\epsilon_{(i)}\}_{i=1}^n$ is ordered, we have $\epsilon_{(i)}\leq\epsilon_{(k)}$ if and only if $i\leq k$, and hence $\sum_{i=1}^n\mbI(i\leq k)=n\tau_k$, which causes the second term in the above equation to vanish. The objective thus simplifies to
\small
\begin{equation}    \label{eq:score_weight}
\mcL^{\rm (GRR)}(\vect{\beta})=\sum_{i=1}^na_n(i) \cdot \epsilon_{(i)}
~\mbox{with}~a_n(i) = \sum_{k=1}^n w_k\big[ \tau_k-\mbI\{ \epsilon_{(i)}\leq\epsilon_{(k)} \} \big] =\sum_{k=1}^n w_k\tau_k-\sum_{k=i}^n w_k.
\end{equation}
\normalsize
When the weights  $\{w_k\}_{k=1}^n$ in  \eqref{eq:wcqr} are non-negative, the score function $a_n(i)$ is non-decreasing with respect to $i$. On the other hand, when the weights $\{w_k\}_{k=1}^n$ take negative values or when $a_n$ is not monotone, both objectives $\mcL^{\rm(GRR)}$ and $\mcL^{\rm(CQR)}$ may become non-convex, thereby extending our framework beyond existing literature. Additionally, in the special case where $w_k\equiv\frac{1}{n(n+1)}$, we establish a connection between equal-weighted CQR and GRR with Wilcoxon score (Example 1), where $a_i=\frac{i}{n+1}-\frac12$ according to \eqref{eq:score_weight}. This derivation also clarifies why these special cases correspond to a $U$-statistic objective, whereas the general cases do not. Finally, deriving the inverse transformation from $a_n(i)$ to $w_i$ is straightforward from \eqref{eq:score_weight}:
\begin{equation}    \label{eq:inverse_trans}
    w_i = a_n(i+1) - a_n(i),\quad i = 1, \ldots ,n-1.
\end{equation}
From the perspective of the score-generating function, \eqref{eq:inverse_trans} suggests setting $w_i$ proportional to the derivative of $\varphi$ at $\tau_i$, that is, $w_i\propto \varphi'(\tau_i)$.
\end{remark}

Remark \ref{rem:connect_cqr} establishes a connection between GRR with general scores and weighted CQR with $K=n$ through the correspondence of $\{a_n(i)\}$ and $\{w_i\}$ in \eqref{eq:score_weight} and \eqref{eq:inverse_trans}. Consequently, one may conjecture that their asymptotic variances mirror each other. This hypothesis is formalized in the following proposition.

\begin{proposition} \label{prop:equiv_jd_mq}
Let $\{w_i\}$ satisfy $|w_i-w_{i+1}|=O(1/K)$ except for a finite number of indices $i$. Let the score function $a_K(i)$ be as in \eqref{eq:score_weight}. Then, the asymptotic variance of GRR, established in Theorem \ref{thm:jd_norm}, is given by $\sigma_{\vect{v}}^2=\lim_{K\rightarrow\infty}V_{K}\bigl(\{\tau_i\},\{w_i\},\vect{v}\bigr)$, where $\tau_i=i/K$ and
\begin{equation}    \label{eq:var_wcqr}
V_{K}\bigl(\{\tau_i\},\{w_i\},\vect{v}\bigr)=\frac{\sum_{i,j=1}^Kw_iw_j \min(\tau_i,\tau_j)\{1-\max(\tau_i,\tau_j)\}}{ [ \sum_{i=1}^Kw_if\{F^{-1}(\tau_i)\} ]^2}  \vect{v}^{\tp}\vect{\Sigma}^{-1}\vect{v}.
\end{equation}
\end{proposition}

Proposition \ref{prop:equiv_jd_mq} shows that the asymptotic variance of weighted CQR, $V_{K}$ for a fixed $K$ \citep{jiang_etal.2012sinica}, converges to that of GRR with the score function given in \eqref{eq:score_weight} as $K\rightarrow\infty$. Intuitively, the numerator and denominator in \eqref{eq:var_wcqr} can be approximated by
\begin{align*}
  \int_0^1\int_0^1\min(u,v)\{1-\max(u,v)\}\,\diff \varphi(u)\diff\varphi(v)=     1 ~~\mbox{ and }~~
    \int_0^1f(F^{-1}(u))\,\diff \varphi(u) 
    = c_H  ,
\end{align*}
based on the correspondence in \eqref{eq:inverse_trans} between the weight in CQR and the score function in GRR, $w_k\propto \diff\varphi(\tau_k)$. 
Detailed calculations are provided in the supplementary material. 
Note that directly determining the optimal weight specification in CQR involves inverting a $K\times K$ matrix, which becomes computationally burdensome and unstable when $K$ is large. This poses a dilemma between statistical efficiency and computational feasibility.
However, by leveraging \eqref{eq:inverse_trans}, 
we can approximate the optimal weight as proportional to the derivative of the optimal score function determined in Proposition \ref{prop:opt_score}. This offers an intuitive and streamlined interpretation of the optimal weight of CQR. Overall, the above derivations establish a clear connection between GRR and weighted CQR, offering both theoretical simplifications and practical insights.

\begin{remark}
The proposed GRR estimator is intimately connected to the classical $M$-estimators \citep{huber1964robust} and $Z$-estimators \citep{godambe.1960aoms}. Specifically, for a univariate loss function $\mcL$ with derivative $\mcL' = \psi$, the $M$-estimator and corresponding $Z$-estimator defined by
\begin{equation*}
    \tilde{\vect{\beta}} = \argmin{\vect{\beta}}\sum_{i=1}^n\mcL(Y_i-\vect{X}_i^{\tp}\vect{\beta}) \quad\text{and}\quad \sum_{i=1}^n\vect{X}_i\psi(Y_i-\vect{X}_i^{\tp}\tilde{\vect{\beta}})=\vect{0},
\end{equation*}
are equivalent and possess the asymptotic variance factor
\begin{equation*}
    \frac{\int_{-\infty}^{\infty}\psi(x)^2f(x)\diff x}{\{\int_{-\infty}^{\infty}\psi'(x)f(x)\diff x\}^2} =  \frac{\int_{-\infty}^{\infty}\psi(x)^2f(x)\diff x}{\{\int_{-\infty}^{\infty}\psi(x)f'(x)\diff x\}^2},
\end{equation*}
where the factor is minimized by the optimal score function $\psi^* = f'/f$. This reveals a direct correspondence between $\psi \circ F^{-1}$ in the $M$-/$Z$-estimation framework and $\varphi$ in our GRR framework in the neighborhood of the model parameter $\vect{\beta}^*$.

Importantly, this correspondence holds only \emph{locally} at $\vect{\beta}^*$. To see this, observe that the GRR loss is asymptotically equivalent to the population-level function
\begin{equation*}
    \sum_{i=1}^n\varphi(F_{\vect{\beta}}(Y_i-\vect{X}_i^{\tp}\vect{\beta}))(Y_i-\vect{X}_i^{\tp}\vect{\beta}),
\end{equation*}
which yields the corresponding estimating equation
\begin{equation*}
    \sum_{i=1}^n\varphi(F_{\vect{\beta}}(Y_i-\vect{X}_i^{\tp}\vect{\beta}))\vect{X}_i = \sum_{i=1}^n(\psi\circ F^{-1})(F_{\vect{\beta}}(Y_i-\vect{X}_i^{\tp}\vect{\beta}))\vect{X}_i = \vect{0}.
\end{equation*}
Here $F_{\vect{\beta}}$ denotes the cumulative distribution function of $Y-\vect{X}^{\tp}\vect{\beta}$, which adapts to the parameter $\vect{\beta}$, whereas the score function $\psi$ in an $M$-/$Z$-estimator is fixed. Therefore, the GRR estimator is \emph{not globally equivalent to} the corresponding $M$-/$Z$-estimator. 
\end{remark}

\begin{remark}
Throughout the paper, we assume that the noise $\epsilon$ is independent of the covariates $\vect{X}$ \citep{zou_yuan.2008aos, wang_peng_etal.2020jasa, zhou_wang_zou.2023jasa, feng_etal.2025arXiv}. Rank regression is designed to robustly estimate the conditional mean while minimizing efficiency loss, which is fundamentally different from single-level quantile regression. The latter targets a specific conditional quantile of the response and therefore permits the more general condition that the $\tau$-th conditional quantile of $\epsilon$ given $\vect{X}$ is zero. In contrast, when efficiency preservation is the primary objective, it is generally difficult to remove the independence assumption. To see this, note that the first-order optimality condition for the GRR objective requires
\begin{equation*}
    \mbE\{ \varphi(F_{\epsilon}(\epsilon))\vect{X}\} = \mbE[\mbE\{ \varphi(F_{\epsilon}(\epsilon) )|\vect{X}\} \vect{X}] = \vect{0},
\end{equation*}
where $F_{\epsilon}$ is the \emph{unconditional} cumulative distribution function of $\epsilon$. Under heteroscedasticity, the conditional distribution of $\epsilon$ given $\vect{X}$ varies with $\vect{X}$, and hence $\mbE \{ \varphi(F_{\epsilon}(\epsilon)) |\vect{X} \}$ need not vanish. Consequently, the GRR estimator $\hat{\vect{\beta}}$ generally converges to a distinct population minimizer, denoted by $\vect{\beta}_0$. Although $\vect{\beta}_0 \neq \vect{\beta}^*$ in general, it remains a meaningful and well-defined population quantity, which can be interpreted as the best linear projection of the heterogeneous rank-based effects onto the covariate space.
\end{remark}

\begin{remark}\label{rem:jiang_compare}
GRR benefits from the efficiency improvements of weighted CQR compared to other QR variants. Conversely, and more importantly, GRR is preferred over weighted CQR both theoretically and practically from several perspectives, as discussed below.

\begin{enumerate}
    \item[(a)] While the limiting variance of the weighted CQR converges to that of GRR when $K=n$, this is practically infeasible as the number of parameters would exceed the sample size. When $K$ is a fixed number, the efficiency loss can be substantial. For example, under a heavy-tailed underlying distribution like $t_{0.1}$, the variance of GRR is only about $85\%$ of that of weighted CQR with $K=20$.  
    
    \item[(b)] From a practical standpoint, weighted CQR requires estimating $K$ parameters $b_{\tau_k}^*$ for $k\hspace{-0.2em}=\hspace{-0.2em}1, \ldots ,K$. This introduces another source of efficiency loss in finite samples. Our simulation studies in Section \ref{sec:sim} further confirm that GRR estimators consistently outperform weighted CQR in terms of both statistical efficiency and computational time.

    \item[(c)] Another challenge for weighted CQR is the potential non-convexity of the loss function due to the presence of negative weights. To date, no fast algorithm has been developed to address weighted CQR with non-convexity.  While linear programming algorithms can help mitigate this issue, they do not scale well with increasing dimensionality. In contrast, we propose a sub-gradient-based algorithm that guarantees fast, scalable, and global convergence for GRR, as demonstrated in Section~\ref{sec:jd_gd} below. 
\end{enumerate}
\end{remark}
\section{{\large A Two-Stage Algorithm for Solving GRR}}
\label{sec:jd_gd}

In the previous section, we present statistical properties of the stationary point $\hat{\vect{\beta}}$ of the GRR objective \eqref{eq:beta_hat_def}.  However, obtaining such stationary points is indeed challenging in practice due to the non-smooth and non-convex nature of the objective with certain score functions. A notable example is illustrated in Figure \ref{fig:nonconvex_loss}: the GRR objective corresponding to the optimal score \eqref{eq:opt_score} under a mixture normal error distribution is both non-convex and non-smooth. To address this challenge, we introduce an iterative algorithm based on sub-gradients and establish the statistical properties of its iterates, which offer greater practical value.

Starting from an initial parameter ${\vect{\beta}}^{(0)}$, at each iteration $t= 0, 1, \ldots$,  a one-step update is performed using the the sub-gradient $\vect{g}_n(\cdot)$ as follows:
\vspace{-0.25em}\begin{equation}\label{eq:subg}
    {\vect{\beta}}^{(t+1)} = {\vect{\beta}}^{(t)} - \eta^{(t)} \vect{g}_n({\vect{\beta}}^{(t)}), ~~\text{ where }~  \vect{g}_n({\vect{\beta}}^{(t)}) = -\frac{1}{n}\sum_{i=1}^na_n(R_i^{(t)})\vect{X}_{i} 
\end{equation}
\vspace{-0.3em}and $\eta^{(t)}>0$ is the step size/learning rate. Here, $R_i^{(t)}$ denotes the rank of $Y_i - \vect{X}_i^{\tp}{\vect{\beta}}^{(t)}$ within the set $\{Y_i-\vect{X}_i^{\tp}{\vect{\beta}}^{(t)}\}_{i=1}^n$. The classical convergence guarantee of sub-gradient descent with a decreasing step size $\eta^{(t)}$ \citep{nesterov2003introductory} is critically applied to convex objectives, failing otherwise. This reliance on convexity poses direct challenges, as the GRR objective \eqref{eq:beta_hat_def} may indeed be non-convex. Moreover, the sub-gradient of non-convex objectives may not be well-defined everywhere. In such cases, the vector $\vect{g}_n(\vect{\beta})$ defined in \eqref{eq:subg} is not necessarily a sub-gradient in the conventional sense but instead belongs to the \emph{Clarke subdifferential} $\partial\mcL_n(\vect{\beta})$, serving as a generalization of ordinary subdifferential sets  \citep{clarke1990optimization}, which enables the analysis of non-convex non-smooth functions using convex analysis techniques.

While the GRR objective \eqref{eq:beta_hat_def} is globally non-convex, we observe that it exhibits local convexity in a neighborhood of the true parameter $\vect{\beta}^*$. Given the distinct landscape properties of the objective—global non-convexity requiring careful initial exploration and local convexity permitting faster refinement—a two-stage approach with tailored step-size schemes for each phase becomes not merely advantageous but essential for achieving robust convergence. Consequently, to effectively address the challenges common to such non-convex loss functions, we advocate for a \emph{two-stage sub-gradient descent} algorithm following this design.

\begin{enumerate}
    \item[(a)] In the first phase, we construct a \emph{convex surrogate loss} for the original objective and apply sub-gradient descent to this surrogate until its iterates enter the neighborhood of the true parameter $\vect{\beta}^*$. For this sub-gradient descent, we employ a carefully chosen decaying step size to ensure global convergence from an arbitrary initialization.
    
    \item[(b)] In the second phase, we revert to the original potentially non-convex loss and apply sub-gradient descent with a constant step size to expedite local convergence.
\end{enumerate}
The innovations and rationales behind this algorithm are detailed in following discussions.

\subsection{Stage one: sub-gradient descent using a convex surrogate}

In stage one, the iterate may be far from the true parameter, leading to optimization in a non-convex region of the target loss. To ensure convergence, it is beneficial to use an alternative convex loss function, such as the GRR objective with the Wilcoxon score. For clarity, we use $\tilde\mcL(\cdot)$ to denote the surrogate loss function optimized during the first stage. The objective $\tilde{\mcL}(\cdot)$ is convex and differs from the original objective $\mcL(\cdot)$ of interest, potentially resulting in stationary points different from each other. Nonetheless, the expectations of the two losses share the same stationary point, according to Theorem \ref{thm:jd_norm}.

\begin{algorithm}[!t]
	\caption{{Early-stage sub-gradient descent for GRR}}
	\label{alg:early_gd_rankr}
	{\textbf{Input:} Dataset $\{(\vect{X}_i,Y_i)\}$, the number of iterations $T$, decaying step sizes sequence $\{\eta^{(t)}\}$.\\ \mbox{}} 	
	\begin{algorithmic}[1]\vspace{-1.1em}
		\STATE Set an initial estimate $\tilde{\vect{\beta}}^{(0)}$. 
		\STATE Specify an arbitrary monotonically increasing score function $\tilde a_n(i)$ in the augmented GRR objective $\tilde{\mcL}(\vect{\beta})$ with form \eqref{eq:beta_hat_def} (e.g., the Wilcoxon scores in Example 1).
		\FOR{$t=1,\dots, T$}
		\STATE Sort the residuals $\{Y_i-\vect{X}_i^{\tp}\tilde{\vect{\beta}}^{(t-1)}\}_{i=1}^n$ to obtain their ranks $\{R^{(t-1)}_i\}_{i=1}^n$.
           
            \STATE Compute the sub-gradient of $\tilde \mcL(\vect{\beta})$ \vspace{-0.7em}\begin{equation*}
                    \tilde{\vect{g}}(\tilde{\vect{\beta}}^{(t-1)}) = -\frac{1}{n}\sum_{i=1}^n\tilde a_n(R^{(t-1)}_i)\vect{X}_{i}.            \end{equation*} 
            \STATE \vspace{-0.75em} Perform sub-gradient descent with diminishing step sizes \begin{equation*}
                    \tilde{\vect{\beta}}^{(t)} = \tilde{\vect{\beta}}^{(t-1)} - \eta^{(t-1)} \tilde{\vect{g}}(\tilde{\vect{\beta}}^{(t-1)}).
                \end{equation*}
            \ENDFOR
            \STATE Compute $\tilde{\vect{\beta}}^{(T,*)} = \argmin{\vect{\beta}\in\{\tilde{\vect{\beta}}^{(t)}\}_{t=0}^T}\tilde \mcL(\vect{\beta})$.
	\end{algorithmic}
	  \textbf{Output:}  The final parameter $\tilde{\vect{\beta}}^{(T,*)}$, serving as the input of the post-stage Algorithm \ref{alg:post_gd_rankr}.
\end{algorithm}

Following the guidelines outlined in Section 3 of \cite{nesterov2003introductory}, we select decaying step sizes $\eta^{(t)}=Ct^{-\zeta}$, $\zeta\in(1/2,1)$ that satisfy the following conditions:
\vspace{-0.5em}\begin{equation}\label{eq:decay_step}
\lim_{T\rightarrow\infty}\sum_{t=1}^T\eta^{(t-1)}=\infty,\quad \lim_{T\rightarrow\infty}\sum_{t=1}^T(\eta^{(t-1)})^2<\infty.
\end{equation}
Importantly, this step size ensures adequate exploration initially without decaying too fast, yet diminishes sufficiently to guarantee eventual convergence,  inherently precluding a constant step size. Upon completing $T$ iterations, we obtain a sequence $\{\tilde{\vect{\beta}}^{(t)}\}_{t=0}^T$ and use\vspace{-0.5em}
\begin{equation}    \label{eq:early_selection}
    \tilde{\vect{\beta}}^{(T,*)} = \argmin{\vect{\beta}\in\{\tilde{\vect{\beta}}^{(t)}\}_{t=0}^T}\tilde{\mcL}(\vect{\beta})
\end{equation}
as the final estimate. The complete algorithm for this stage is provided in Algorithm \ref{alg:early_gd_rankr}. We can establish the convergence of the final estimator from this stage as $|\tilde{\vect{\beta}}^{(T,*)} - \hat{\vect{\beta}}|_2=O(T^{(\zeta-1)/2})$, $\zeta\in(1/2,1)$, based on Theorem S1 and the local strong convexity of the GRR loss. 
This essentially indicates that the iterates of Algorithm \ref{alg:early_gd_rankr} will enter the neighborhood of $\vect{\beta}^*$ as $T$ grows, and can therefore yield a statistically consistent estimator. We leave a complete theoretical statement of the convergence guarantee to Section C of the supplementary material. 
However, Algorithm \ref{alg:early_gd_rankr} has limitations:
\begin{enumerate}
    \item[(a)] First, to guarantee global convergence, the surrogate loss $\tilde\mcL(\cdot)$ should be convex and may preclude the direct use of the loss corresponding to optimal score functions (e.g., Example 4). In such instances, the output estimate  will not be statistically efficient. 
    \item[(b)] Second, the decaying step size $\eta^{(t)} = Ct^{-\zeta}$ satisfying \eqref{eq:decay_step} leads to a suboptimal (slow) algorithmic convergence rate $O(t^{(\zeta-1)/2})$, $\zeta\in(1/2,1)$, as $t$ iterates. 
\end{enumerate}These factors motivate a second algorithmic stage designed to accelerate both algorithmic convergence and statistical efficiency.

\subsection{Stage two: sub-gradient descent with the non-convex loss}
\begin{algorithm}[!t]
	\caption{{ Post-stage sub-gradient descent for GRR}}
	\label{alg:post_gd_rankr}
    \textbf{Input:} The output of early-stage Algorithm \ref{alg:early_gd_rankr},  $\tilde{\vect{\beta}}^{(T,*)}$, dataset $\{(\vect{X}_i,Y_i)\}$, the number of iterations $T$,  constant step size $\eta$.\\ \mbox{} 	
	\vspace{-1.2em}\begin{algorithmic}[1]
    \STATE Set the initial parameter $\hat{\vect{\beta}}^{(0)}=\tilde{\vect{\beta}}^{(T,*)}$ as the output of Algorithm \ref{alg:early_gd_rankr}.
    \FOR{$t=1,\dots, T$}
        
		\STATE Sort the residual $\{Y_i-\vect{X}_i^{\tp}\hat{\vect{\beta}}^{(t-1)}\}_{i=1}^n$ and obtain the ranks $\{R^{(t-1)}_i\}_{i=1}^n$. 
            \STATE Compute the (Clarke) sub-gradient of $\mcL(\vect{\beta})$:
                \vspace{-0.75em}\begin{equation*}
                    \vect{g}\big(\hat{\vect{\beta}}^{(t-1)}\big) = -\frac{1}{n}\sum_{i=1}^na_n(R^{(t-1)}_i)\vect{X}_{i},
                \vspace{-.7em}\end{equation*}
                using $a_n(\cdot)$, the score function in GRR \eqref{eq:beta_hat_def} that can be potentially non-monotonic. 
            \STATE Perform sub-gradient descent with a constant step size $\eta$
                \begin{equation*}
                    \hat{\vect{\beta}}^{(t)} = \hat{\vect{\beta}}^{(t-1)} - \eta \vect{g}\big(\hat{\vect{\beta}}^{(t-1)}\big).
                \vspace{-.85em}\end{equation*}
            \ENDFOR
	\end{algorithmic}
    \textbf{Output:}  The final parameter $\hat{\vect{\beta}}^{(T)}$.
\end{algorithm}

To expedite convergence and achieve statistical efficiency, we perform sub-gradient descent on the original, potentially non-convex, loss function $\mcL(\vect{\beta})$ during the second stage. Distinct from the first stage, this phase employs a constant step size, and the final iterate $\hat{\vect{\beta}}^{(T)}$ serves as the estimator. Algorithm~\ref{alg:post_gd_rankr} details the complete procedure. We establish its convergence rate in Theorem S2 and Corollary S2 of the supplementary material. These results show that the algorithmic error in the population landscape converges geometrically, at a rate determined by the step size $\eta$, until the overall error is dominated by the statistical remainder term $R_{n,T}$, which depends on the continuity of the score-generating function $\varphi$. Importantly, Corollary S2 shows that, under proper initialization and a sufficient number of iterations, the output of our two-step algorithm, $\hat{\vect{\beta}}^{(T)}$, has the \emph{same} asymptotic distribution as the stationary point $\hat{\vect{\beta}}$ established in Section \ref{sec:grr}. This is notable because $\hat{\vect{\beta}}^{(T)}$ may \emph{not} be identical to the theoretical stationary point $\hat{\vect{\beta}}$.

\begin{remark}  \label{rem:dev_exact}
    In classical settings where the loss function is both smooth and strongly convex, gradient descent algorithms with constant steps achieve geometric convergence to the empirical minimizer \citep{nesterov2003introductory}. Our objective however is both non-smooth and non-convex, rendering existing theoretical results and proof techniques inapplicable. Consequently, departing from standard practice, we do not establish the convergence of iterates towards the empirical stationary point. Instead, we demonstrate the convergence towards the underlying model parameter $\vect{\beta}^*$, albeit with a non-vanishing remainder term. This shift in analytical focus is crucial for leveraging the strong convexity of the population risk in the vicinity of $\vect{\beta}^*$.
\end{remark}

\begin{remark}[Phase Transition in Algorithmic Convergence] The distinct convergence behavior (c.f., Theorems S1 and S2) characterize an algorithmic phase transition between the two stages of the algorithm. Specifically, in stage one, the iterations converge at a polynomial rate $O(t^{(\zeta-1)/2})$. In stage two, the algorithm further converges at a geometric rate of $O(\|\mbI - \eta c_H\vect{\Sigma}\|^{t})$ towards the optimal statistical error identified by the main term $-c_H^{-1}\vect{\Sigma}^{-1}\vect{h}(\vect{\beta}^*)$. In summary, given an arbitrary initialization, it may take a polynomial number of steps in stage one to enter a benign region in the vicinity of $\vect{\beta}^*$, where the population loss is strongly convex (even though the empirical loss may not yet share this property). Subsequently, only an additional $O(\log n)$ iterations are needed in stage two due to the geometric convergence. This overall two-stage strategy achieves a statistically efficient solution with a compellingly small number of iterations for this challenging non-smooth, non-convex problem. 
\end{remark}

\section{{\large Statistical Inference via Multiplier Bootstrap}}
\label{sec:bootstrap}

As demonstrated in Theorem \ref{thm:jd_norm}, the asymptotic variance of $\hat{\vect{\beta}}$ depends on a constant $c_H$ specified in Condition \ref{cond:score}, which poses challenges for direct estimation. Therefore, the bootstrap method naturally arises as a preferred approach for constructing confidence intervals. 
In this section, we explore the use of multiplier bootstrap \citep{spokoiny_zhilova.2015} to conduct inference on the GRR estimator.
Let $B\geq 1$ denote the number of bootstrap samples. For each $1\leq b\leq B$, we independently generate $n$ Rademacher random variables $\{e_{i,b}\}_{i=1}^n$, satisfying $\mbP(e_{i,b}=1) = \mbP(e_{i,b}=-1) = 1/2$, and minimize the weighted objective to obtain the bootstrapped GRR estimator
\begin{equation}    \label{eq:MB_loss}
    \hat{\vect{\beta}}^{(b)} = \argmin{\vect{\beta} \in \mathbb{R}^p}\mcL^b(\vect{\beta}) = \argmin{\vect{\beta} \in \mathbb{R}^p } \frac{1}{n}\sum_{i=1}^n(1+e_{i,b})a_n(R_i)(Y_i - \vect{\beta}^{\tp}\vect{X}_i) .
\end{equation}
Subsequently, confidence intervals can be constructed based on the bootstrapped estimates $\{\hat{\vect{\beta}}^{(b)}\}_{b=1}^B$. However, due to the non-smoothness and non-convexity of the loss functions $\mcL(\vect{\beta})$ and $\mcL^b(\vect{\beta})$, solving \eqref{eq:MB_loss} exactly for $B$ times can be computationally prohibitive, especially for large $B$. Instead, drawing inspiration from Theorem S2, we can initialize with an arbitrary estimate $\hat{\vect{\beta}}^{(0)}$ that falls into the region of geometric convergence, and then apply sub-gradient descent with a constant step size for the bootstrapped loss, ensuring geometric convergence.

More specifically, given an initial estimate $\hat{\vect{\beta}}^{(0)}$ in round $b$, we generate $n$ i.i.d. random weights $\{1+e_{i,b}\}_{i=1}^n$ and obtain gradient descent iterates 
     $\hat{\vect{\beta}}^{(t,b)} = \hat{\vect{\beta}}^{(t-1,b)} - \eta \vect{g}^b(\hat{\vect{\beta}}^{(t-1,b)})$, 
where $\vect{g}^b(\hat{\vect{\beta}}^{(t-1,b)})$ denotes the sub-gradient of the bootstrapped loss \eqref{eq:MB_loss}, that is,
\begin{equation*}
    \vect{g}^b(\hat{\vect{\beta}}^{(t-1,b)}) = -\frac{1}{n}\sum_{i=1}^n(1+e_{i,b})a_n(R_i^{(t-1,b)})\vect{X}_{i},\quad 1\leq t\leq T.
\end{equation*}
After $T$ iterations, for each $1\leq l\leq p$, we construct the $100(1-\alpha)$\% confidence interval $[z^b_l(\alpha/2),z^b_l(1-\alpha/2)]$ for $\beta^*_l$, where $\alpha\in(0,1)$ and
\begin{equation*}
    \begin{aligned}
        z^b_l(\alpha/2) = &\inf\{z\in\mbR:\mbP^*(\hat{\beta}_{l}^{(T,b)}\leq z)\geq \alpha/2\},\\
        z^b_l(1-\alpha/2) = &\inf\{z\in\mbR:\mbP^*(\hat{\beta}_{l}^{(T,b)}\leq z)\geq 1-\alpha/2\} .
    \end{aligned}
\end{equation*}
The detailed steps are outlined in Algorithm~\ref{alg:rankr_gd_MB}. Specifically, we compute the initial estimate by an early-stopped gradient descent with a convex GRR loss outlined in Algorithm~\ref{alg:early_gd_rankr}. To conclude this section, we present the theoretical justification for the validity of the multiplier bootstrap method in approximating the distribution of the GRR estimator.

\begin{algorithm}[!t]
	\caption{{ Multiplier bootstrap of GRR}}
	\label{alg:rankr_gd_MB}
	\hspace*{\algorithmicindent}   {\textbf{Input:} Dataset $\{(\vect{X}_i,Y_i)\}$, number of iterations $T$, step size $\eta$. \\ \mbox{}} 	
	\begin{algorithmic}[1]
	\vspace{-1.3em}	\STATE Compute an initial parameter $\hat{\vect{\beta}}^{(0)}$ by solving \eqref{eq:beta_hat_def} with Algorithm \ref{alg:early_gd_rankr}. 
            
		\FOR{$b=1,2,\dots,B$}
                \STATE Generate $n$ i.i.d. random weights $\{1+e_{i,b}\}_{i=1}^n$ satisfying $\mbP(e_{i,b}=1)=\mbP(e_{i,b}=-1)=\frac12$.
                \vspace{-0.9em} \FOR{$t=1,2,\dots, T$}                
		      \STATE Sort the residual $\{Y_i-\vect{X}_i^{\tp}\hat{\vect{\beta}}^{(t-1,b)}\}_{i=1}^n$ and obtain the ranks $\{R^{(t-1, b)}_i\}_{i=1}^n$.
                \STATE Compute the sub-gradient
                    \vspace{-0.75em}\begin{equation*}
                        \vect{g}^b(\hat{\vect{\beta}}^{(t-1,b)}) = -\frac{1}{n}\sum_{i=1}^n(1+e_{i,b})a_n(R^{(t-1,b)}_i)\vect{X}_{i}.   \vspace{-1.3em} \end{equation*}
                \STATE Perform sub-gradient descent 
                    \vspace{-0.3em} \begin{equation*}
                        \hat{\vect{\beta}}^{(t,b)} = \hat{\vect{\beta}}^{(t-1,b)} - \eta \vect{g}^b(\hat{\vect{\beta}}^{(t-1,b)}).
                    \vspace{-0.8em} \end{equation*}
                \ENDFOR
            \ENDFOR
            \FOR{$l = 1,2,\dots,p$}
                \STATE For the $l$-th coordinate, compute the $\alpha/2$- and $(1-\alpha/2)$-quantile by\vspace{-0.3em} \begin{equation*}
                \begin{aligned}
                    z^b_l(\alpha/2) = &\inf\{z\in\mbR:\mbP^*(\hat{\beta}_{l}^{(T,b)}\leq z)\geq \alpha/2\},\\
                    z^b_l(1-\alpha/2) = &\inf\{z\in\mbR:\mbP^*(\hat{\beta}_{l}^{(T,b)}\leq z)\geq 1-\alpha/2\}.\vspace{-0.3em} 
                \end{aligned}
                \end{equation*}
            \ENDFOR
	\end{algorithmic}
	 \textbf{Output:} The $p$-tuple confidence intervals $\bigl\{[z^b_l(\alpha/2),z^b_l(1-\alpha/2)]\bigr\}_{l=1}^p$.
\end{algorithm}

\begin{theorem} \label{thm:MB_prob}
 Assume that Conditions \ref{cond:cov}--\ref{cond:score} hold. Let the initial estimate $\hat{\vect{\beta}}^{(0)}$ satisfy $|\hat{\vect{\beta}}^{(0)}-\vect{\beta}^*|_2\leq C_0\sqrt{p\log(n)/n}$ with probability tending to one, for some constant $C_0>0$. Set a constant step size $\eta$ such that $\|\mbI - \eta c_H\vect{\Sigma}\|<1$.
 Let $T \geq C_1 \log n$ for sufficiently large $C_1$. When $p^3=o(n/(\log n)^3)$, for any $\vect{v} \in \mbS^{p-1}$, 
    \begin{equation*}
\sup_{x\in\mbR}\Big|\mbP\big\{ \sqrt{n}\bigl\langle\hat{\vect{\beta}}^{(T)} - \vect{\beta}^*,\vect{v}\bigr\rangle\leq x\big\} - \mbP^*\big\{ \sqrt{n}\bigl\langle\hat{\vect{\beta}}^{(T,b)} - \hat{\vect{\beta}}^{(T)},\vect{v}\bigr\rangle\leq x\big\} \Big|\xrightarrow{\mbP}0,\quad \text{as }n \rightarrow \infty,
    \end{equation*}
    where $\mbP^*$ denotes the conditional probability, given the observed dataset $\{(\vect{X}_i,Y_i)\}_{i=1}^n$, and $\vect{v}$ refers to a sequence $\vect{v}_n$ of unit vectors in $\mbR^p$ with p possibly growing with n.
\end{theorem}

Theorem \ref{thm:MB_prob} guarantees that the empirical distribution of $\hat{\vect{\beta}}^{(T,b)} - \hat{\vect{\beta}}$ well approximates the distribution of $\hat{\vect{\beta}} - \vect{\beta}^*$. The condition on the initial estimator can be easily satisfied by the two-stage algorithm proposed in Section \ref{sec:jd_gd}. On the other hand, the condition can be weakened at the cost of additional terms in the Bahadur remainder, for which the detailed representation is presented in Theorem S3, Section C of the supplementary material.


\section{{\large Numerical and Empirical Studies}}
\label{sec:sim}

In this section, we examine the empirical performance of GRR and its comparison to alternatives, using simulated datasets as well as a real dataset on bike sharing demand. More simulation results are presented in Section A of the supplementary materials.


\subsection{Simulation studies}	\label{sec:syn}

In our simulation studies, the data vectors $\{ (Y_i, \vect{X}_i \}_{i=1}^n$ are generated from the 
linear model $Y = \vect{X}^{\tp}\vect{\beta}^*+\epsilon$, where $\vect{\beta}^* = (1, \dots,1)^{\tp} \in \mathbb{R}^p$, $\vect{X}=\vect{\Sigma}^{1/2}\tilde{\vect{X}}$ with $\vect{\Sigma} = (0.7^{|j-k|})_{1\leq j, k\leq p }$  and $\tilde{\vect{X}}$ containing independent entries drawn from Uniform$[-\sqrt{3/2},\sqrt{3/2}]$. The noise variable $\epsilon$ follows one of the following three distributions:
\begin{itemize}
    \item Cauchy: The standard Cauchy distribution;   
    \item Gaussian mixture: A location mixture normal given by $0.5 \mcN(-\frac{3}{2},\frac{1}{100})+ 0.5 \mcN(\frac{3}{2},\frac{1}{100})$;
    \item Smoothed uniform: The distribution $U+ 0.1\,Z$, where $U\sim \mathrm{Unif}[-1,1]$ and $Z\sim\mcN(0,1)$ are independent.
\end{itemize}
\noindent
We compare the four candidate methods (SRR, WRR, ASM, and ORR) against oracle benchmarks in our numerical experiments:
\begin{itemize}
    \item SRR: Single-level generalized rank regression (Example 3) with  $\tau = 1/2$;
    \item WRR: Wilcoxon score-based rank regression (Example 1), as a special case of GRR;
    \item ASM: The antitonic score matching estimator proposed in \cite{feng_etal.2025arXiv};
    \item \textbf{ORR(est)}: Optimal score-based rank regression with the score function estimated via three-fold cross-fitting \citep{feng_etal.2025arXiv};
    \item \textbf{ORR(orc)}: Optimal score-based rank regression with an oracle optimal score function computed explicitly from the known noise distribution;
    \item Oracle MLE: The maximum likelihood estimator computed under the exact true noise distribution. Since this estimator is infeasible in practice, it serves as a theoretical lower bound for the estimation error.
\end{itemize}
We estimate the optimal score function $f'(F^{-1}(u))/f(F^{-1}(u))$ using kernel-based estimators for both the density and its derivative. Following \cite{schuster1969estimation}, we adopt a bandwidth equal to twice Silverman's rule-of-thumb bandwidth for estimating the density derivative. To mitigate numerical instability in the tails, where probability masses become small, we compute the derivative of $f(F^{-1}(u))$ with respect to $u$ directly, rather than estimating the derivative separately. This implementation aligns with that of \cite{feng_etal.2025arXiv}. Detailed mathematical formulations for constructing the optimal score function are provided in the supplementary material; see Appendix H.

\begin{table}[H]
\centering
\caption{The $\ell_2$ estimation errors and their corresponding standard errors (in parentheses). The noise term $\epsilon$ is generated from a $\mathrm{Cauchy}(0,1)$ distribution, a Gaussian mixture distribution, and a smoothed uniform distribution. We compare four candidate estimators (SRR, WRR, ASM, and ORR) against the oracle benchmark MLE, which assumes perfect knowledge of the underlying noise distribution.}
\label{tab:orr_error}
\resizebox{16cm}{6.5cm}{
\begin{tabular}{cccccccc}
\toprule
\multirow{2}{*}{Noise} & \multirow{2}{*}{Method} & \multicolumn{2}{c}{$n=1800$} & \multicolumn{2}{c}{$n=2400$} & \multicolumn{2}{c}{$n=3000$} \\
\cmidrule(r){3-4} \cmidrule(r){5-6} \cmidrule(r){7-8} & & $p=5$ & $p=10$ & $p=5$ & $p=10$ & $p=5$ & $p=10$\\
\midrule
\multirow{6}{*}{Cauchy(0,1)} 
& SRR & 0.157(0.057) & 0.232(0.054) & 0.134(0.053) & 0.201(0.059) & 0.123(0.044) & 0.174(0.048) \\
& WRR & 0.156(0.053) & 0.218(0.048) & 0.132(0.043) & 0.189(0.049) & 0.125(0.043) & 0.165(0.043) \\
& ASM & 0.175(0.069) & 0.261(0.063) & 0.148(0.062) & 0.221(0.069) & 0.131(0.047) & 0.193(0.057) \\
& \textbf{ORR(est)} & 0.125(0.039) & 0.165(0.038) & 0.104(0.036) & 0.145(0.036) & 0.100(0.034) & 0.129(0.032) \\
& \textbf{ORR(orc)} & 0.126(0.039) & 0.168(0.038) & 0.103(0.038) & 0.150(0.040) & 0.099(0.032) & 0.131(0.033) \\
& Oracle MLE & 0.123(0.039) & 0.165(0.035) & 0.101(0.037) & 0.145(0.038) & 0.099(0.031) & 0.127(0.032) \\
\hline 
\multirow{6}{*}{\makecell{$\frac{1}{2}\mcN(-\frac{3}{2},\frac{1}{100})$\\$+\frac{1}{2}\mcN(\frac{3}{2},\frac{1}{100})$}} 
& SRR & 0.481(0.101) & 0.546(0.093) & 0.455(0.090) & 0.508(0.087) & 0.447(0.084) & 0.501(0.093) \\
& WRR & 0.024(0.008) & 0.036(0.011) & 0.021(0.008) & 0.031(0.010) & 0.018(0.008) & 0.028(0.008) \\
& ASM & 0.015(0.006) & 0.023(0.006) & 0.014(0.005) & 0.020(0.005) & 0.012(0.005) & 0.018(0.005) \\
& \textbf{ORR(est)} & 0.013(0.005) & 0.019(0.005) & 0.012(0.005) & 0.017(0.005) & 0.011(0.004) & 0.018(0.004) \\
& \textbf{ORR(orc)} & 0.012(0.005) & 0.018(0.005) & 0.011(0.004) & 0.015(0.004) & 0.009(0.004) & 0.014(0.003) \\
& Oracle MLE & 0.011(0.004) & 0.016(0.004) & 0.010(0.004) & 0.014(0.004) & 0.008(0.003) & 0.013(0.003) \\
\hline
\multirow{6}{*}{\makecell{ Smoothed\\Uniform } }
& SRR & 0.090(0.030) & 0.132(0.038) & 0.074(0.027) & 0.114(0.028) & 0.073(0.024) & 0.104(0.027) \\
& WRR & 0.056(0.020) & 0.086(0.022) & 0.050(0.017) & 0.072(0.016) & 0.049(0.016) & 0.065(0.015) \\
& ASM & 0.039(0.016) & 0.060(0.014) & 0.034(0.015) & 0.050(0.012) & 0.031(0.012) & 0.045(0.011) \\
& \textbf{ORR(est)} & 0.042(0.014) & 0.057(0.014) & 0.038(0.013) & 0.047(0.012) & 0.035(0.011) & 0.043(0.010) \\
& \textbf{ORR(orc)} & 0.038(0.013) & 0.052(0.013) & 0.035(0.013) & 0.043(0.011) & 0.032(0.011) & 0.040(0.009) \\
& Oracle MLE & 0.034(0.013) & 0.051(0.013) & 0.031(0.012) & 0.043(0.011) & 0.028(0.010) & 0.039(0.009) \\
\bottomrule
\end{tabular}
}
\end{table}

The comprehensive results are reported in Table \ref{tab:orr_error}. We make the following observations.
\begin{itemize}
    \item Across all three noise settings, ORR(est) achieves $\ell_2$ estimation errors that are remarkably close to those of the Oracle MLE. It is important to note that the Oracle MLE, included here as an absolute benchmark, assumes perfect knowledge of the underlying noise distribution, which is generally unavailable in practice. Notably, the fully data-driven ORR(est) attains estimation accuracy that is nearly indistinguishable from this theoretical lower bound across all settings, demonstrating the near-optimal performance of the proposed estimator.    
    \item On average, ORR(est) reduces the $\ell_2$ estimation error by approximately $31\%$ relative to WRR and by approximately $14\%$ relative to ASM. The proposed ORR(est) method is comparable to, and in severely non-convex scenarios consistently outperforms, the ASM method \citep{feng_etal.2025arXiv}. The latter provides an elegant solution by identifying the best approximation within the restricted class of convex losses, thereby avoiding the need to design tailored algorithms for non-convex optimization landscapes. In contrast, our GRR framework relaxes this convexity constraint and directly targets the exact optimal score function. By addressing the resulting non-convexity through a theoretically justified two-stage algorithm, GRR achieves higher statistical efficiency in settings where the optimal landscape is non-convex.
\end{itemize}
The strong empirical performance of ORR(est) highlights the practical reliability of our two-stage algorithm in solving non-convex objectives, supported by the theoretical guarantees established in Theorems S1 and S2 of the supplementary material.

\subsection{A real data example}	\label{sec:real_data}

In this section, we evaluate our methodology on the Seoul Bike Sharing Demand dataset\footnote{https://archive.ics.uci.edu/dataset/560/seoul+bike+sharing+demand}. The target variable is the hourly rented bike count, with standard deviation $645$, and the covariates include temperature, humidity, wind speed, visibility, and other weather and seasonal variables. In this real-data example, the primary goal is not to recover a true underlying regression parameter, since the linear model may be misspecified and the error distribution may exhibit heterogeneity. Instead, we focus on a practically relevant comparison of predictive accuracy and inferential efficiency, as measured by out-of-sample prediction error and confidence interval width. In particular, we use this example to examine whether adaptive score estimation can improve finite-sample performance relative to standard rank-based alternatives.

The dataset consists of $n=8,760$ samples with dimension $p=12$. Prior to our analysis, we center and normalize both the feature vectors $\vect{X}$ and the response $Y$. We randomly split the data into a training set of $8,100$ samples and a testing set of $660$ samples. Since the true noise distribution is unknown, we compare SRR, WRR, and $\mathrm{ORR(est)}$.

For the optimization algorithm, we use $50$ initial iterations with a decaying step size $\eta^{(t)} = t^{-\frac{2}{3}}$, together with a normalization of the initial gradient update to stabilize the early stage of training. The algorithm then switches to a constant step size $\eta = 50^{-\frac{2}{3}}$ for the remaining $150$ iterations. To assess the stability of the results, we compute the average prediction error on the testing set over $100$ independent random data splits. We also report the corresponding confidence interval widths. To ensure that the interval widths are sufficiently stable, we use 1000 bootstrap iterations in the inference stage. \\
\begin{figure}[htp]
\begin{subfigure}[t]{0.45\textwidth}
  \includegraphics[width=\linewidth]{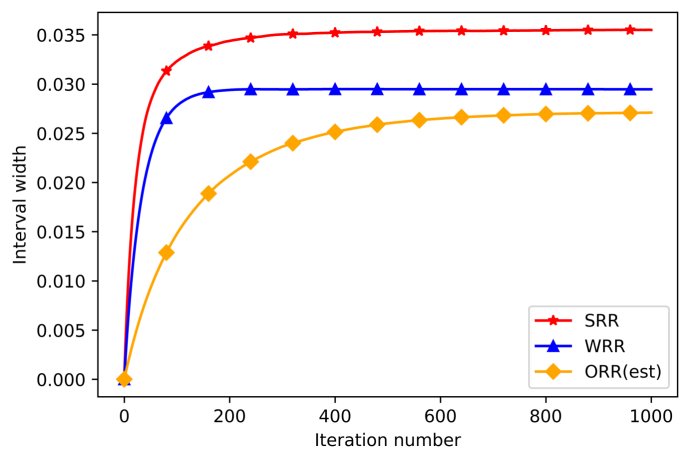}
\end{subfigure}\hfill
\begin{subfigure}[t]{0.45\textwidth}
    \includegraphics[width=\linewidth]{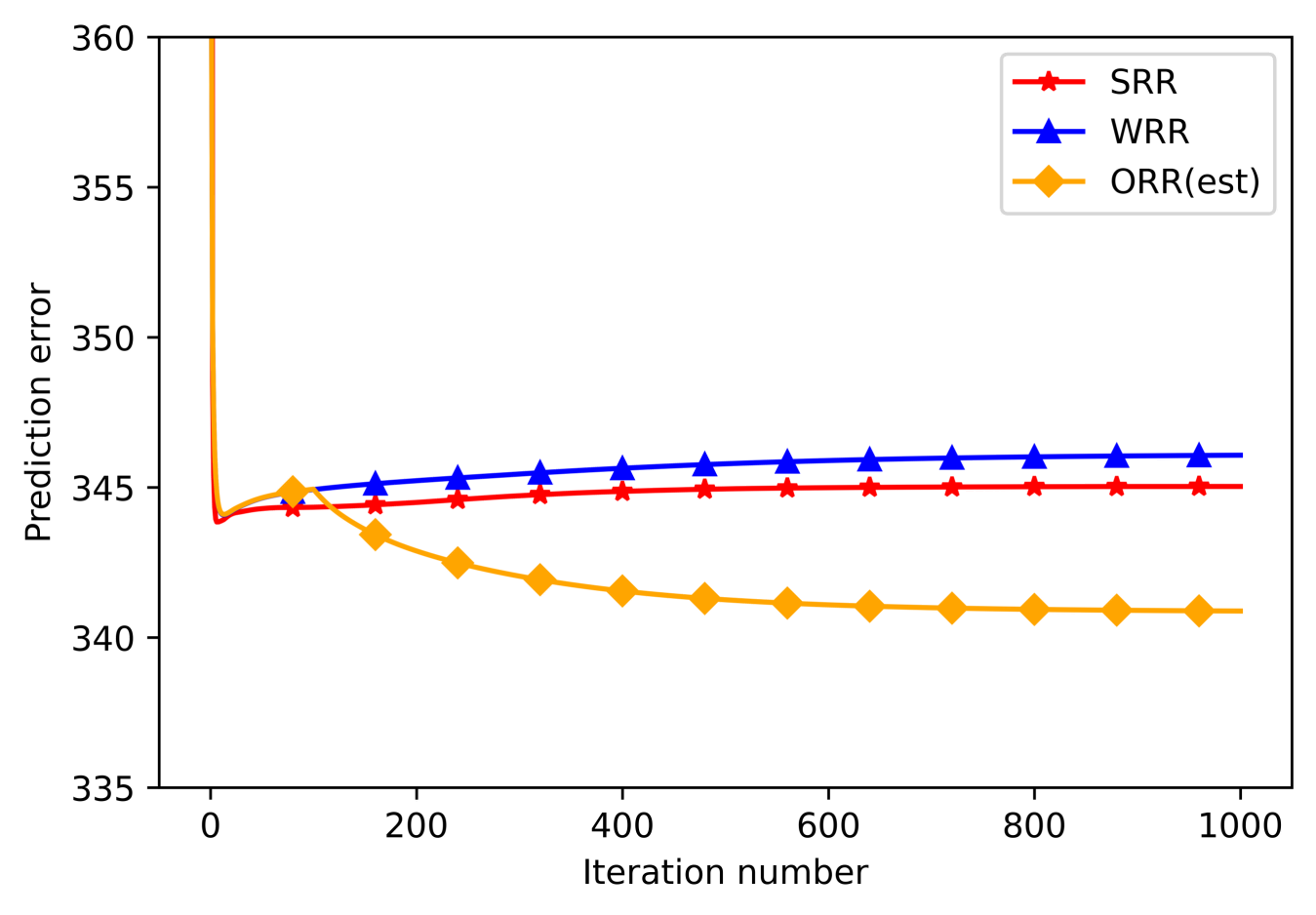}
\end{subfigure}\hfill
\caption{Confidence interval width (first column) and prediction error (second column) versus the number of iterations for the methods evaluated on the Seoul Bike Sharing Demand dataset.}
\label{fig:realdata_iter}
\end{figure}
Figure \ref{fig:realdata_iter} shows that ORR(est) consistently outperforms SRR and WRR in terms of both confidence interval width and prediction error. Although SRR achieves slightly lower prediction error than WRR, its bootstrapped confidence intervals are notably wider. These results suggest that adaptively estimating and incorporating distributional information into the GRR score function can improve both predictive and inferential performance.


\section{{\large Concluding Remarks}}
\label{sec:conclude}
This paper introduces a generalized rank regression framework that improves statistical efficiency and robustness over canonical rank regression by directly employing the exact score function tailored to the underlying error distribution. Our work presents four primary contributions. First, we establish the non-asymptotic theory for GRR, explicitly addressing the non-smooth and non-convex loss landscapes, thereby achieving efficiency gains over methods restricted to convex approximations. Second, we propose a tailored two-stage sub-gradient descent algorithm with theoretical guarantees of global convergence to the statistically efficient solution within a small number of iterations. Third, we introduce a variant of the multiplier bootstrap for conducting reliable statistical inference. Additionally, as an interesting byproduct, we reveal close theoretical connections between GRR under certain score functions and variants of quantile regression, which, in turn, explains the advantages of GRR in terms of statistical efficiency. The GRR framework is further extended to high-dimensional sparse models in this work,
while a comprehensive systematic study of this extension is deferred to future research.

\bibliographystyle{asa}
\bibliography{GRR_arXiv}

\end{document}